\documentclass[12pt]{iopart}

\usepackage{graphicx}
\newcommand{\PSAW}{P_\mathrm{SAW}}				% SAW power
\newcommand{\vSAW}{v_\mathrm{SAW}}				% SAW velocity
\newcommand{\lSAW}{\lambda_\mathrm{SAW}}		% SAW wavelength
\newcommand{\fSAW}{f_\mathrm{SAW}}				% SAW frequency
\newcommand{\VSAW}{\phi_\mathrm{SAW}}			% SAW piezoelectric potential
\newcommand{\Prf}{P_\mathrm{rf}}				% rf-Power

\newcommand{\keff}{k_{eff}^2}					% acousto-electric coupling coefficient

\newcommand{\sr}[1]{s_\mathrm{#1#1}}			% reflexion coefficient
\newcommand{\st}[0]{s_\mathrm{21}}				% transmission coefficient, s21
\newcommand{\Iae}[0]{I_\mathrm{ae}}				% acousto-electric current
\newcommand{\sxx}[0]{\sigma_{xx}}					% longitudinal conductivity
\newcommand{\sM}[0]{\sigma_\mathrm{M}}			% characteristic conductivity
\newcommand{\Vg}[0]{V_\mathrm{gate}}				% gate voltage

\begin{document}

\topical{Interaction of surface acoustic waves with electronic excitations in graphene}
\author{A. Hern\'andez-M\'inguez, Y.-T. Liou and P. V. Santos}
\address{Paul-Drude-Institut f\"ur Festk\"orperelektronik, Leibniz-Institut im Forschungsverbund Berlin e.V., Hausvogteiplatz 5-7, 10117 Berlin, Germany}
\ead{alberto.h.minguez@pdi-berlin.de}
\vspace{10pt}
%\begin{indented}
%\item[]\today
%\end{indented}

\begin{abstract}
This article reviews the main theoretical and experimental advances regarding the interaction between surface acoustic waves (SAWs) and electronic excitations in graphene. The coupling of the graphene electron gas to the SAW piezoelectric field can modify the propagation properties of the SAW, and even amplify the intensity of SAWs traveling along the graphene layer. Conversely, the periodic electric and strain fields of the SAW can be used to modify the graphene Dirac cone and to couple light into graphene plasmons. Finally, SAWs can generate acousto-electric currents in graphene. These increase linearly with the SAW frequency and power but, in contrast to conventional currents, they depend non-monotonously on the graphene electric conductivity. Most of these functionalities have been reported in graphene transferred to the surface of strong piezoelectric insulators. The recent observation of acousto-electric currents in epitaxial graphene on SiC opens the way to the large-scale fabrication of graphene-based acousto-electric devices patterned directly on a semi-insulating wafer.
\end{abstract}

%\pacs{}
\vspace{2pc}
%\noindent{\it Keywords\/}: graphene, surface acoustic waves, acousto-electric interaction

%\submitto{\JPD}
\maketitle

%\begin{figure}
%\includegraphics[width=0.80.7\linewidth]{}
%\caption{}
%\label{}
%\end{figure}

\tableofcontents

\newpage
%%%%%%%%%%%%%%%%%%%%%%%%%%%%%%%%%%%%%%%%%%%%%%%%%%
\section{Introduction}\label{sec_Intro}

In the last decades, the application of dynamic strain and electric fields of surface acoustic waves (SAWs) has proven to be a successful tool for industrial applications, as demonstrated by the wide spectrum of SAW-based signal processing devices incorporated in electronic systems currently available in the market~\cite{Campbell_98}. Fortunately, this success has not prevented the scientific community from further investigating additional functionalities that could become the grounds of new, future SAW-based applications. As an example, it has been demonstrated that SAWs can manipulate optically generated electronic excitations in low-dimensional semiconductor heterostructures~\cite{Rocke97a, Kinzel_NL11_1512_11, PVS152, PVS234, PVS265, PVS223}. Conversely, SAWs have been used for the dynamic modulation of the non-classical light emitted by semiconductor quantum dots~\cite{Gell_APL93_081115_08, Metcalfe_PRL105_37401_10, PVS218, PVS246, Weiss_APL109_33105_16, Villa_APL111_11103_17}. If the semiconductor heterostructure contains a two-dimensional electron gas, the coupling of SAWs to the charge carriers can also be used to induce electric currents~\cite{Esslinger_SSC84_939_92, Shilton_PRB51_14770_95, Shilton_JoPCM7_7675_95, Rotter_APL73_2128_98}. Moreover, promising functionalities for quantum processing have been demonstrated including the acoustic transport of single electrons along one-dimensional channels~\cite{TSPS97a}, as well as on-demand electron and spin transfer between electrostatic quantum dots~\cite{Hermelin_N477_435_11, McNeil_N477_439_11}. Other potential applications of SAWs that are being investigated are the acoustic modulation of photonic structures~\cite{PVS156}, the manipulation of magnetic states in ferromagnetic materials~\cite{Davis_APL97_232507_10}, and even the coupling of SAWs to superconducting qubits~\cite{Gustafsson_S346_207_14}.

In this context, it is not at all surpising the increasing attention that the interaction between SAWs and graphene has attracted in the last years. Since its discovery in 2004~\cite{Novoselov_S306_666_04}, the peculiar mechanical and electronic properties of this pure two-dimensional material have suggested many promising applications in areas like flexible electronics, biological engineering, composite materials and even in optoelectronics and photovoltaics~\cite{Ferrari_N7_4598_15}. Therefore, it is reasonable to expect that the combination of the functionalities provided by SAWs with the unique electronic properties of graphene will bring forth a new generation of acousto-electric devices for a wide range of technological applications.

The objective of this manuscript is to overview the main advances reported until now about the interaction between SAWs and charge carriers in graphene. The review is organized as follows. Sections~\ref{sec_Graphene}  and ~\ref{sec_SAWs} discuss the most relevant characteristics of the graphene electronic properties and surface acoustic waves, respectively. Section~\ref{sec_SAWmodification} addresses the effects of the SAW-graphene coupling on the propagation of the SAWs, while Section~\ref{sec_Graphenemodulation} focuses on the use of the SAW periodic potentials for the modification of the graphene electronic properties and the coupling of light to graphene electronic excitations. Finally, Section~\ref{sec_AEcurrents} is devoted to the SAW-induced generation of acousto-electric currents in graphene. We have grouped the results reported here in two subsections, depending on whether the acousto-electric current is generated in graphene transferred to a piezoelectric substrate (Section~\ref{sec_transferred}), or in epitaxial graphene formed on SiC (Section~\ref{sec_epitaxial}). Finally, Section~\ref{sec_Concl} summarizes the main conclusions of this work.

\section{Basic properties of graphene and SAWs}\label{sec_MainProperties}

\subsection{Graphene}\label{sec_Graphene}

\begin{figure}
\centering
\includegraphics[width=0.6\linewidth]{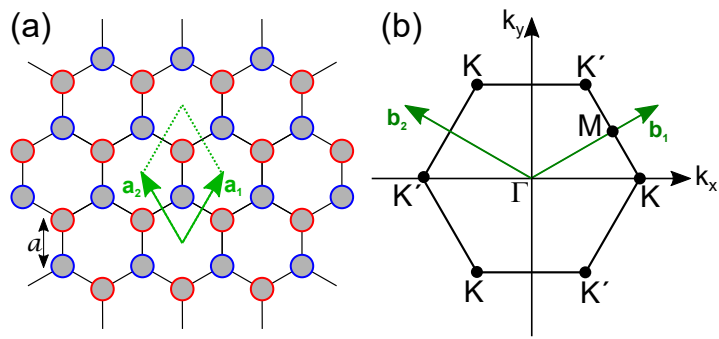}
\caption{(a) Honeycomb lattice structure of graphene with unit vectors $\mathbf{a_1}$ and $\mathbf{a_2}$. The sites of the two non-equivalent triangular sub-lattices are marked as red and blue circumferences. The distance between two neighbor atoms is $a=1.42$~\AA. (b) First Brillouin zone of the reciprocal lattice with unit vectors $\vec{b}_1$ and $\vec{b}_2$. The Dirac points are located at the non-equivalent $K$ and $K'$ points at the corners of the Brillouin zone.}
\label{fig_Graphene}
\end{figure}

Graphene consists of a single layer of carbon atoms forming an hexagonal lattice, as shown in Fig.~\ref{fig_Graphene}(a). This atomic arrangement is described by a triangular lattice with unit vectors $\vec{a}_1$ and $\vec{a}_2$ (throughout the manuscript, over-arrows will represent vectors and bold letters will indicate tensors) and a basis of two atoms per unit cell. The hexagonal lattice can also be seen as the superposition of two non-equivalent triangular sub-lattices marked as red and blue circumferences in Fig.~\ref{fig_Graphene}(a), where each pair of nearest-neighbor atoms occupy sites belonging to different sub-lattices. Figure~\ref{fig_Graphene}(b) displays the first Brillouin zone of the graphene reciprocal lattice, together with its reciprocal unit vectors $\vec{b}_1$ and $\vec{b}_2$. The reciprocal lattice is also hexagonal and, due to the two non-equivalent sub-lattices in the real space, the corners are also grouped in two non-equivalent $K$ and $K'$ points. 

%%%%%%%%%%%%%%%%%%%%%%%%%%%%%%%%%%%%%%%%%%%%%%%%%%%%%%%%%%
%%%%%%%%%%%%%%%%%%%%%%%%%%%%%%%%%%%%%%%%%%%%%%%%%%%%%%%%%%

The electronic band structure of graphene is well-described theoretically by a tight-binding model~\cite{Wallace_PR71_622_47}. Here, we will just present its most relevant features, as a comprehensive discussion can be found elsewhere~\cite{CastroNeto_RMP81_109_09}. Taking only electron hopping between nearest-neighbor atoms into account, the energy dispersion is expressed as:

\begin{equation}\label{eq_DispRelation1}
E(k_x,k_y)=\pm\gamma_0\left[3 + 2\cos(\sqrt{3}k_xa) + 4\cos(\frac{3}{2}k_ya)\cos(\frac{\sqrt{3}}{2}k_xa)\right]^{1/2}.
\end{equation}

\begin{figure}
\centering
\includegraphics[width=0.6\linewidth]{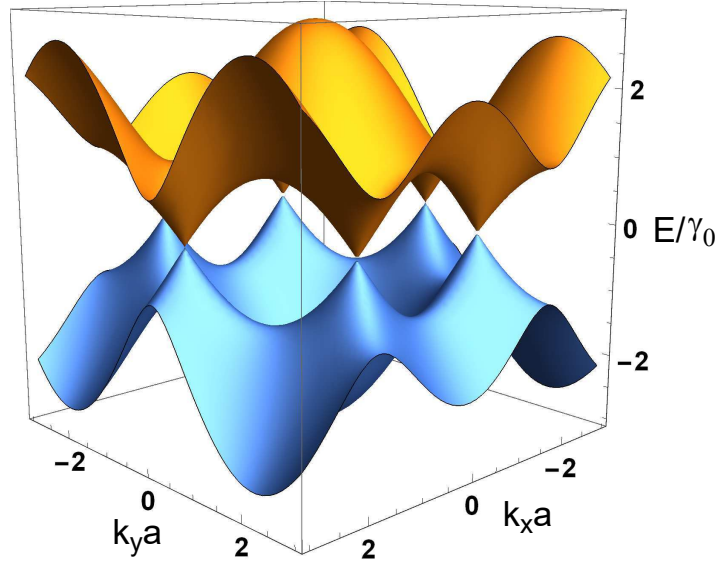}
\caption{Electronic dispersion relation of graphene calculated by taking only into account hopping between nearest neighbor atoms. The dispersion relation is linear around the Dirac points.}
\label{fig_BandStruct}
\end{figure}

\noindent Here, $\vec{k}=(k_x,k_y)$ is the crystal momentum of the graphene charge carriers, $\gamma_0=2.8$~eV is the nearest-neighbor hopping energy, and the distance between two nearest-neighbor atoms is $a=1.42$~\AA. Figure~\ref{fig_BandStruct} displays the electronic band structure given by Eq.~\ref{eq_DispRelation1}. It consists of a valence band and a conduction band touching each other at the six $K$ and $K'$ points of the first Brillouin zone. This means that graphene has no band gap between valence and conduction band, and therefore it behaves electrically as a semi-metal. In electrically neutral graphene, the valence band is fully occupied, so that the Fermi energy, $E_F$, intersects the electronic bands at the $K$ and $K'$ points, also known as charge-neutrality points or Dirac points.

At low energies around the charge-neutrality points, the electronic dispersion of Eq.~\ref{eq_DispRelation1} can be approximated as:

\begin{equation}\label{eq_DispRelation2}
E(\vec{k})\approx\pm\hbar v_F|\vec{k}|.
\end{equation}

\noindent Here, $\hbar$ is the reduced Planck constant, $v_F=3\gamma_0a/2\hbar\approx10^6$~m/s is the Fermi velocity, and $\vec{k}$ is expressed relatively to the wave vector at the charge-neutrality point. According to Eq.~\ref{eq_DispRelation2}, the energy dispersion around the charge-neutrality point is not parabolic, but linear, like the energy dispersion of massless relativistic Dirac fermions in quantum electrodynamics. Therefore, many of the specific properties of Dirac fermions are also valid for charge carries in graphene, with the peculiarity that, in the last case, the electrons move with an effective speed $v_F$ which is 300 times slower than the speed of light. 
 
The linear energy dispersion of graphene has important consequences on its electronic properties. For example, the density of states of graphene is not independent of the energy level, like in conventional 2D electron gases, but linear according to:

\begin{equation}
D(E)=\frac{2A_c}{\pi}\frac{|E|}{\hbar^2v_F^2},
\end{equation}

\noindent where $A_c=3\sqrt{3}a^2/2$ is the area of the unit cell. Another consequence is that the presence of a magnetic field $B$ perpendicular to the graphene surface does not lead to equally spaced Landau energy levels, but to relativistic Landau levels at energies $E(n)=\pm v_F\sqrt{2e\hbar B|n|}$ with $n\in\mathbf{Z}$, and an additional level at the charge-neutrality point.

%%%%%%%%%%%%%%%%%%%%%%%%%%%%%%%%%%%%%%%%%%%%%%%%%%%%%%%%%%
%%%%%%%%%%%%%%%%%%%%%%%%%%%%%%%%%%%%%%%%%%%%%%%%%%%%%%%%%%

Due to the presence of the two non-equivalent sub-lattices in the atomic structure of graphene, the electron wave function near the charge-neutrality point is expressed as a two-component spinor, where each component indicates the relative contribution of each sub-lattice to the electronic state. This introduces an additional degree of freedom named ``pseudospin'' due to its analogy with the electronic spin. A relevant quantity to characterize the graphene electronic wave functions is their helicity or chirality, defined as the projection of the pseudospin operator along de direction of motion. Electrons and holes in graphene have a positive and negative helicity, respectively, which implies that electrons with wave vector $\vec{k}$ and holes moving along the  $-\vec{k}$ direction are intricately connected. The well-defined chirality of the charge carriers is an important property because many interesting electronic phenomena in graphene are based on the conservation of this quantity, e.g. Klein tunneling~\cite{Katsnelson_NP2_620_06}.   

The most ``simple'' way to obtain graphene is mechanical exfoliation of bulk graphite~\cite{Novoselov_S306_04}. This technique provides nowadays high-quality graphene layers up to 100 $\mu$m size. Although this is sufficient for research purposes, commercial applications require the reproducible fabrication of large-area graphene layers at a relatively low cost. Among the different techniques that have been explored to grow graphene epitaxially, the two more promising ones are based on chemical vapor deposition (CVD) and on thermal decomposition of SiC. In the first case, monolayer graphene is formed on metal surfaces by catalytic decomposition of hydrocarbons or carbon oxide \cite{Yu_APL93_113103_08, Li_S324_1312_09}. In the second case, silicon atoms desorb from the top layers of SiC upon heating, leaving graphene layers on the surface~\cite{Berger_JPCB108_19912_04, Berger_S312_1191_06}. The quality and number of such layers depends on the SiC face on which graphene is formed, as well as on the temperature and duration of the heating treatment.

%%%%%%%%%%%%%%%%%%%%%%%%%%%%%%%%%%%%%%%%%%%%%%%%%%%%%%%%%%%%%%%%%%%%%%%
\subsection{Surface acoustic waves}\label{sec_SAWs}

%%%%%%%%%%%%%%%%%%%%%%%%%%%%%%%%%%%%%%%%%%%%%%%%%%%%%%%%%%
\begin{figure}
\centering
\includegraphics[width=0.6\linewidth]{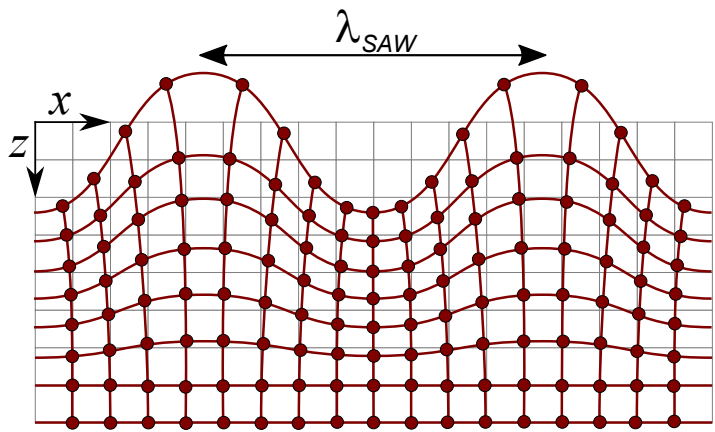}
\caption{Representation of a surface Rayleigh wave propagating along the $\hat{x}$ axis. It consists of the superposition of a longitudinal acoustic (LA) and a transverse acoustic mode polarized perpendicular to the surface (TA$_z$). The reference system is such that the half-space $z>0$ fills the elastic medium.}
\label{fig_SAW}
\end{figure}
%%%%%%%%%%%%%%%%%%%%%%%%%%%%%%%%%%%%%%%%%%%%%%%%%%%%%%%%%%

SAWs are low-frequency acoustic phonons confined to the surface of an unbounded solid~\cite{LL65a,Auld90a}. They result from special solutions of the elastic wave equation for the acoustic displacement field, $\vec{u}(\vec{r},t)$, near a surface:

\begin{equation}\label{eq_ElasticWave}
\vec{\nabla}\cdot\mathbf{T}=\rho\frac{\partial^2\vec{u}}{\partial t^2},
\end{equation}

\noindent where $\mathbf{T} $ is the stress tensor and $\rho$ is the density of the medium. If the solid is a piezoelectric material, then the solutions of the elastic equation must also satisfy the constitutive relations:

\begin{equation}\label{eq_ConstRelations1}
\mathbf{T}=\mathbf{c}\mathbf{S}-\mathbf{e}\vec{E},
\end{equation}

\begin{equation}\label{eq_ConstRelations2}
\vec{D}=\mathbf{e}\mathbf{S}+\mathbf{\varepsilon}\vec{E}.
\end{equation}

\noindent Here, $\mathbf{S}$ is the strain tensor, $\mathbf{c}$ the elastic stiffness coefficient, $\mathbf{\varepsilon}$ the dielectric constant, $\mathbf{e}$ the piezoelectric coefficient, $\vec{D}$ the electric displacement field and $\vec{E}=-\vec{\nabla}\VSAW$ is the piezoelectric field, which can be expressed in terms of the piezoelectric potential, $\VSAW$. As the acoustic propagation velocity is about $10^5$ times slower than the velocity of light, the values of $\mathbf{D}$ and $\mathbf{E}$ at each given time can be we well approximated in many cases by their static solutions (quasi-static approximation)~\cite{Auld90a,Royer00a}, where the Maxwell equations are replaced by the condition $\vec{\nabla}\cdot\vec{D}=0$ if free charge carriers are not present in the material. In a stress-free surface, SAW modes are obtained by looking for solutions of Eqs.~\ref{eq_ElasticWave}, \ref{eq_ConstRelations1} and \ref{eq_ConstRelations2} that decay towards zero when $z$ penetrates into the bulk (throughout this manuscript, we will use the coordinate system defined in Fig.~\ref{fig_SAW}, where the elastic medium fills the $z>0$ half-space, and the SAW propagates along the $x$-direction) and also satisfy the following boundary conditions at $z=0$:

\begin{eqnarray}
T_{iz}=0, \\
D_z(z=0^+)=D_z(z=0^-),
\end{eqnarray}

\noindent Note that, contrary to $\vec{u}$, $\VSAW$ also extends into the vacuum side ($z<0$), where it decays with the same characteristic length as into the bulk.

Among the different kinds of SAWs, Rayleigh modes~\cite{Rayleigh_PLMSs1-17_4_85} are of particular importance. They consist of the superposition of a longitudinal acoustic mode (LA) with a transverse acoustic mode polarized perpendicular to the surface (TA$_z$), cf. Fig.~\ref{fig_SAW}. For weak piezoelectric materials, analytical solutions of Rayleigh SAWs can be derived for both $\vec{u}$ and $\VSAW$ (see, for instace, Ref.~\cite{Simon_PRB54_13878_96}). The situation becomes more complex when the SAW propagates along a layered medium. Here, the continuity conditions for $T_{iz}$ and $D_z$, as well as for the SAW velocity, $\vSAW$, must be satisfied not only at the top surface, but also across each interface plane~\cite{TAST98a}. In this case, surface solutions are usually obtained by solving the coupled elastic and electric equations numerically. 

In piezoelectric materials, SAWs can be electrically generated using an inter-digital transducer (IDT)~\cite{White_APL7_314_65}. This consists of a metal grating patterned on the surface of the material, where the fingers are alternatingly connected to an rf-voltage source and the electric ground. The periodicity of the grating determines the wavelength of the SAW generated by the IDT, $\lSAW$. When the frequency of the rf-voltage source, $f$, matches the condition $f=\vSAW/\lSAW$, then the inverse piezoelectric effect transforms a fraction of the electromagnetic energy applied to the IDT into a SAW. The efficiency of the energy conversion depends on the design of the IDT and on the electro-mechanical coupling coefficient of the piezoelectric material, $\keff$~\cite{Campbell_98}. Table~\ref{tab_substrates} summarizes the value of $\keff$ and $\vSAW$ for some of the materials discussed in this review.

%%%%%%%%%%%%%%%%%%%%%%%%%%%%%%%%%%%%%%%%%%%%%%%%%%%%%%%%%%
\begin{table}
\centering
\caption{Electro-mechanical coupling coefficient, $\keff$, and SAW propagation velocity, $\vSAW$, of several materials used to couple SAWs and graphene.}\label{tab_substrates}
\begin{tabular}{lcccc}
\hline 
Material & Crystal & SAW & $\keff$ & $\vSAW$\\ 
 • & cut & direction & (\%) & (m/s) \\ 
\hline 
h-SiC & $<$0001$>$ & $\{1\overline{1}00\}$ & 0.0111 & 6832 \\ 
h-ZnO & $<$0001$>$ & $\{1\overline{1}00\}$ & 1.12 & 2691 \\
LiTaO$_3$ & 36$^\circ$-Y & X & 5.0 & 4160 \\
LiNbO$_3$ & 128$^\circ$-Y & X & 5.3 & 3992 \\

\hline 
\end{tabular}
\end{table}
%%%%%%%%%%%%%%%%%%%%%%%%%%%%%%%%%%%%%%%%%%%%%%%%%%%%%%%%%%

\begin{figure}
\centering
\includegraphics[width=0.6\linewidth]{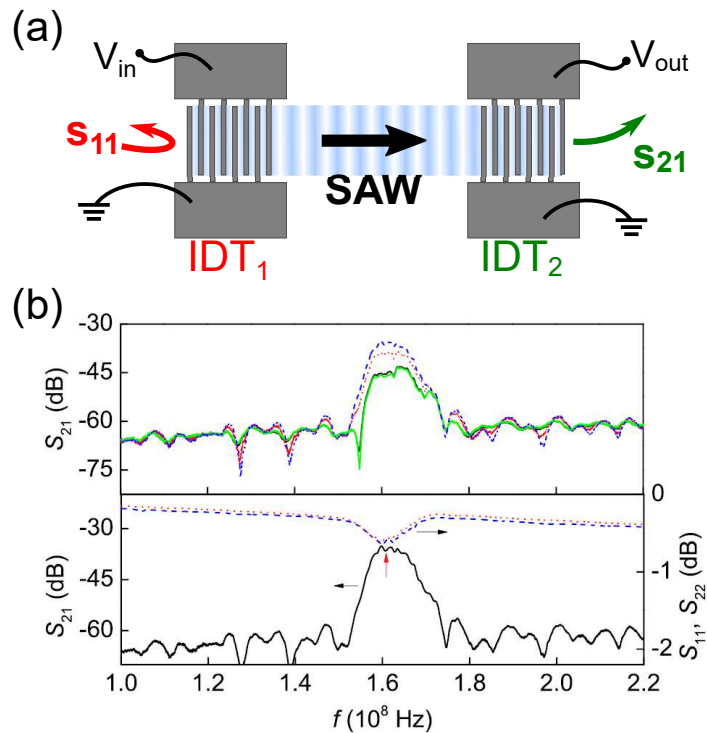}
\caption{(a) SAW delay line consisting of two interdigital transducers (IDTs) patterned on a piezoelectric substrate. Under resonant frequency conditions of the rf-voltage $V_\mathrm{in}$ applied to IDT$_1$, a fraction of the rf-power is transformed into a SAW and travels towards IDT$_2$, where it is partially transformed back into an rf-voltage, $V_\mathrm{out}$.  (b) s-parameters as a function of rf-frequency for a SAW delay line consisting of IDT fingers made of graphene. Upper panel: $\st$ coefficients for as-deposited graphene (black solid line), after exposition to nitric acid vapor during 20~min (blue dashed line), after just 5~min exposition (red dotted line) and 24~h later (green solid line). Lower panel: $\sr1$, $\sr2$ and $\st$ for graphene IDT fingers exposed to nitric acid vapor during 20~min. Reprinted from~\cite{Mayorov_APL104_83509_14} with the permission of AIP Publishing.}
\label{fig_delayline}
\end{figure}

In addition to SAW generation, the IDT can also convert SAWs back into rf-voltages. Many electronic systems take advantage of this property and use SAW delay lines as filters of rf-signals. A delay line consists of two IDTs with the same grating periodicity, placed in front of each other on a piezoelectric substrate, cf. Figure~\ref{fig_delayline}(a). When a multi-frequency rf-signal, $V_\mathrm{in}$, is applied to IDT$_1$, only the frequency component matching the resonant frequency of the transducer is transformed into a SAW. Then, the acoustic wave travels towards IDT$_2$, which transforms the SAW back into an rf-signal of the same frequency, $V_\mathrm{out}$. To determine the insertion loss of a SAW delay line, the most common method is to measure its rf-power reflection and transmission coefficients, $\sr1$ and $\st$ respectively, as a function of the frequency of the rf-signal applied to the input IDT.

%%%%%%%%%%%%%%%%%%%%%%%%%%%%%%%%%%%%%%%%%%%%%%%%%%%%%%%%%%
%%%%%%%%%%%%%%%%%%%%%%%%%%%%%%%%%%%%%%%%%%%%%%%%%%%%%%%%%%

Some of the issues limiting the efficiency of SAW delay lines, especially at large frequencies, are related to the impact of the IDT metal fingers on SAW generation and propagation~\cite{Campbell_98}. From this point of view, IDTs made of graphene could be a promising alternative, because its exceptional lightness is expected to minimize any mass-loading effects. Mayorov \etal~\cite{Mayorov_APL104_83509_14} have reported the first SAW delay line made of graphene IDTs. After transferring CVD graphene to the surface of a strong piezoelectric substrate (LiNbO$_3$ in this case), they patterned it into IDT fingers working at a frequency $\sim164$~MHz. Figure~\ref{fig_delayline}(b) shows the rf-spectra of the $\st$ coefficient for such a device, measured at room temperature. The insertion-loss was -45~dB (upper panel, black solid line), a value 17~dB worse than the insertion-loss of the same delay line made of Cr/Au instead of graphene. The authors attributed this difference to a significant ohmic dissipation in the graphene fingers caused by their relatively large electric resistance, thus reducing their efficiency for SAW generation and detection~\cite{Lakin1974}.

To study the effect of graphene resistance on the efficiency of the IDTs, the authors exposed the SAW delay line to nitric acid vapor. The interchange of electric charges between graphene and the adsorbed molecules leads to and increase of the graphene carrier concentration~\cite{Bae2010}, and therefore to a reduction of the IDT finger resistance. After 20~min exposition to the dopant, the insertion-loss improved by 10~dB (upper panel in Fig.~\ref{fig_delayline}(b), blue dashed line). Interestingly, the performance of graphene IDTs exposed only 5~min to the vapor (red dotted line) reverted back to its original state after a period of 24~h (green solid line) due to evaporation of the dopant from the graphene area. This behavior suggests that graphene IDTs could be used as a new kind of SAW-based light or gas sensors. In conventional designs of such kind of sensors, the graphene layer is placed between the two IDTs forming the delay line~\cite{Arsat_CPL467_344_09, Ciplys_IEEESEnsors_785_10, Chivukula_IToUFaFC59_265_12, Whitehead_APL103_63110_13}, and the change in the response of the delay line is associated to the modifications that the absorption of photons or molecules by the graphene film (e.g. charge doping, mass loading) induce in the propagation properties of the SAW. The response of SAW sensors based on graphene IDTs, in contrast, is based on the modulation that light and/or gas exposition induces on the generation and detection efficiency of the IDTs, thus allowing for smaller, more compact designs. 
 
%%%%%%%%%%%%%%%%%%%%%%%%%%%%%%%%%%%%%%%%%%%%%%%%%%
\section{Interaction of graphene electronic excitations with SAWs}\label{sec_SAW-G}

\subsection{SAW attenuation and amplification}\label{sec_SAWmodification}

It is well known that acoustic vibrations can interact strongly with moving charges present in a piezoelectric semiconductor by means of the strain-induced piezoelectric field~\cite{Parmenter_PR89_990_53, Weinreich_PR104_321_56}. This effect, known as acousto-electric (AE) coupling, is also important when SAWs propagate along a semiconductor that contains a two-dimensional electron gas (2DEG) close to its surface~\cite{Ingebrigtsen_JAP41_454_70, Simon_PRB54_13878_96}. As the energy of the SAW phonons is typically much lower than the band gap energy of the semiconductor, SAWs are usually absorbed by the electron gas via intra-band electronic transitions (i.e. within the conduction or the valence band). If the conductive layer consists of graphene, then its gapless electronic energy dispersion could, in principle, allow SAW absorption by both intra- and inter-band electronic transitions. This possibility, however, has been excluded by Zhang \etal~\cite{Zhang_AA1_22146_11}, because the linear electronic dispersion of graphene makes impossible for any inter-band transition to satisfy both the energy and momentum conservation laws required in such a scattering process.

From the acoustic point of view, the interaction between SAWs and moving charges results usually in the re-normalization of the SAW velocity and in the attenuation of the SAW intensity as it propagates along the 2DEG. This is because the kinetic energy acquired by the charges via SAW absorption is typically lost due to carrier scattering. According to the classical relaxation theory, the velocity shift, $\Delta v$, and the attenuation coefficient, $\alpha$, depend on the longitudinal conductivity of the 2DEG, $\sxx$, as~\cite{Ingebrigtsen_JAP41_454_70, Wixforth_PRL56_2104_86, Simon_PRB54_13878_96}:

\begin{equation}\label{eq_SAWobservables}
\frac{\Delta v}{v_0}-i\frac{\alpha}{q}=\frac{\keff/2}{1+i\sxx(q,\omega,B,\mu)/\sM}.
\end{equation}

\noindent Here, $\keff$ is the electro-mechanical coupling introduced in Table~\ref{tab_substrates}, $\sM$ is a characteristic conductivity that depends on the materials surrounding the 2DEG, and the velocity shift $\Delta v=\vSAW-v_0$ is defined with respect to the SAW velocity $v_0$ when $\sxx\rightarrow\infty$. In general, $\sxx$ is a complex number that depends not only on the chemical potential of the electron gas, $\mu$, and on the presence of an external magnetic field, $B$, but also on the wave vector of the SAW, $q$, and its angular frequency, $\omega=\vSAW q$~\cite{Pan_NatComm8_1243_2017}.

The effect of $\mu$ and $B$ on SAW propagation has been thoroughly studied in e.g. GaAs-based 2DEGs~\cite{Wixforth_PRL56_2104_86, Wixforth89a}. This behavior is expected to be distinctly different in the case of graphene because of the linear dependence of the electronic density of states and the non-equispaced energy distribution of the relativistic Landau levels. Thalmeier \etal~\cite{Thalmeier_PRB81_41409_10} have theoretically studied the dependence of $\Delta v$ and $\alpha$ on graphene carrier density under an external magnetic field. The problem can be substantially simplified by just considering the static regime $\sxx(B,\mu)$. This limit is valid under the assumption  that \textsl{(i)} the typical frequency of SAWs is extremely small compared to the typical electron-hole excitation energy, and \textsl{(ii)} the graphene film contains scattering centers with a scattering rate exceeding $v_Fq$, so that the dependence of $\sxx$ on $q$ can be neglected. Under these conditions, the Shubnikov-de Haas oscillations predicted for $\sxx$ as a function of the chemical potential~\cite{Dora_PRB76_35402_07} will also influence the SAW propagation properties. Every time that $\mu$ crosses a Landau level, the increase of $\sxx$ will induce a dip in $\Delta v$ and, therefore, a decrease in the SAW velocity, while the positioning of $\mu$ between two Landau levels will cause a local maximum of $\Delta v$. Similar oscillations are expected for the SAW attenuation coefficient, with the difference that the crossing of a Landau level will induce either a peak or a dip on $\alpha$ depending on whether $\sxx/\sM\ll1$ or $\sxx/\sM\gg1$, respectively.

%%%%%%%%%%%%%%%%%%%%%%%%%%%%%%%%%%%%%%%%%%%%%%%%%%%%%%%%%%
\begin{figure}
\centering
\includegraphics[width=0.6\linewidth]{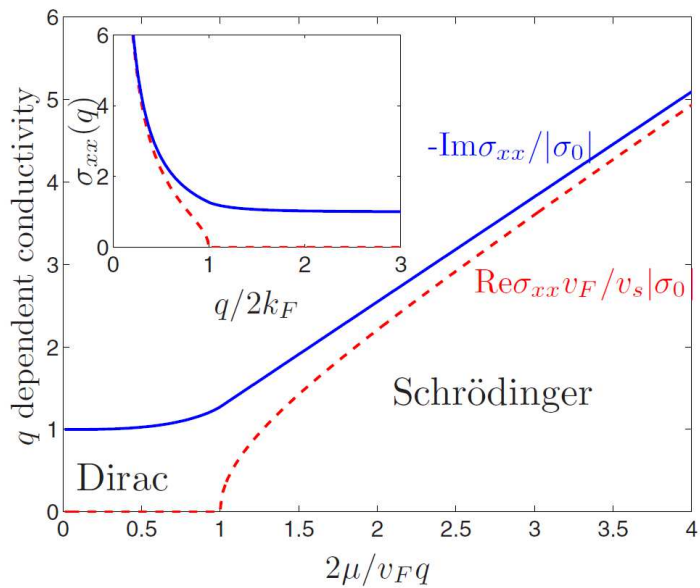}
\caption{Real (red dashed line) and imaginary (blue solid line) part of $\sxx$ as a function of the chemical potential, $\mu$. The Dirac and Schr\"odinger-like regions are separated at $2\mu/v_Fq=1$, and $|\sigma_0|=e^2\vSAW/v_F$. The inset shows $\sxx$ vs. $q$ for a fixed value of the Fermi momentum $k_F=\mu/v_F$. Reprinted figure with permission from~\cite{Thalmeier_PRB81_41409_10}. Copyright (2018) by the American Physical Society.}
\label{fig_Thalmeier}
\end{figure}
%%%%%%%%%%%%%%%%%%%%%%%%%%%%%%%%%%%%%%%%%%%%%%%%%%%%%%%%%%

A case of special interest is the interaction of SAWs with an ultra-clean graphene sheet in the absence of an external magnetic field. Under these conditions, the carrier scattering rate is now smaller than $v_Fq$, and therefore the SAW observables will also provide information about the relation $\sxx(q)$. Figure~\ref{fig_Thalmeier} displays the real and imaginary parts of $\sxx(q,\mu)$ (red dashed line and blue solid line, respectively) calculated for this limit case. There are two distinctive regimes depending on the ratio between $q$ and the Fermi momentum of the charge carriers, $k_F=\mu/v_F$ (expressed in units where $\hbar=1$). For $2k_F>q$, the conductivity follows the typical dependence of a Schr\"odinger metal. The Dirac nature of the graphene electron gas manifests itself when the Fermi momentum is tuned into the region $2k_F<q$. Here, $Re(\sxx)=0$ and $Im(\sxx)=\-e^2\vSAW/v_F$ at the charge-neutrality point. The fact that the conductivity is a pure imaginary magnitude in this regime implies that, according to Eq.~\ref{eq_SAWobservables}, a SAW traveling along the graphene layer will experience renormalization of its velocity, but its intensity will not be attenuated at all.

\begin{figure}
\centering
\includegraphics[width=0.6\linewidth]{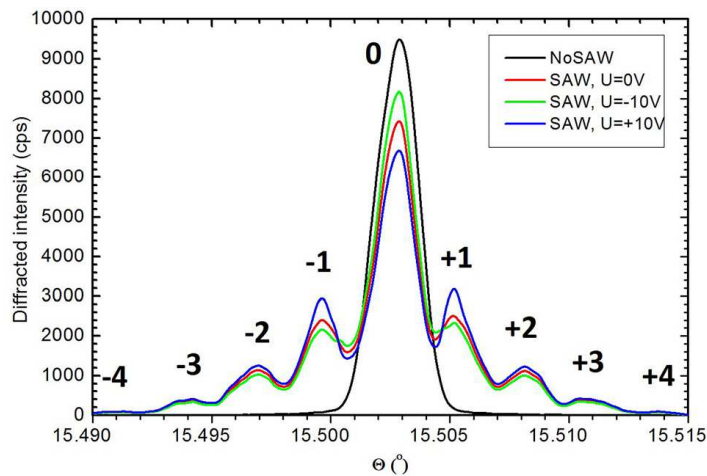}
\caption{Rocking curves of a CTGS crystal modulated by a SAW of $\lSAW=6~\mu$m. In absence of SAW, a single peak is observed (black curve). When the SAW is applied, four diffraction satellites appear at each side of Bragg's peak (red curve). The amplitude of the satellites increases (decreases) under positive (negative) DC voltage applied along the graphene sheet, as a consequence of SAW amplification (attenuation). Reprinted from~\cite{Insepov_APL106_23505_15} with the permission of AIP Publishing.}
\label{fig_Insepov}
\end{figure}

Under special conditions, the interaction between SAW and graphene charge carriers can induce not only SAW attenuation, but also SAW amplification. The amplification of ultrasonic waves was first observed in the 60s in systems combining piezoelectric and semiconductor materials~\cite{Hutson_PRL7_237_61, Collins_APL13_314_68}. It bases on the fact that, if the carriers move under an electric field in the same direction as the SAW with a drift velocity ($v_d$) exceeding $\vSAW$, then the carriers will transfer part of their kinetic energy to the wave~\cite{White_JAP33_2547_62, Gulyaev_SPJ20_1508_65,Pustovoit_SPU12_105_69}. Recently, Insepov \etal have reported SAW amplification in a Ca$_3$TaGa$_3$Si$_2$O$_{14}$ (CTGS) piezoelectric crystal containing a p-doped graphene film at the surface~\cite{Insepov_APL106_23505_15}. In their experiment, the authors visualized SAWs of 471~MHz frequency propagating along the graphene film using an X-ray diffraction method~\cite{Roshchupkin_APA114_1105_14, Roshchupkin_JAP118_104901_15}. Figure~\ref{fig_Insepov} displays several rocking curves of the CTGS crystal obtained by the X-ray diffractometer. In the absence of SAWs, the rocking curve consists of a single Bragg peak (black curve). When an rf-signal of the appropriate frequency was applied to the IDT, then the Bragg peak reduced its amplitude and four diffraction satellites appeared at each side due to the modulation of the crystal lattice by the SAW (red curve). The angular divergence between the satellites is determined by the ratio between the inter-atomic plane distance and the wavelength of the SAW. The application of a DC electric potential, $U$, along the graphene layer modified the relative amplitude of the central and satellite peaks. When $U=-10$~V, the drift velocity of the holes in the graphene film is opposed to $\vSAW$. As a consequence, the SAW transferred part of its energy to the holes, an effect that manifested in the rocking curve as a reduction of the satellite peaks and an enhancement of the central Bragg peak (green curve). The application of $U=+10$~V, in contrast, induced a flow of holes moving in the same direction as the acoustic wave, with $v_d$ exceeding $\vSAW$. Therefore, SAW amplification turned on, as manifested by the increase of the satellite peaks of Fig.~\ref{fig_Insepov} (blue curve) and the further suppression of the central peak with respect to its amplitude at $U=0$.

%%%%%%%%%%%%%%%%%%%%%%%%%%%%%%%%%%%%%%%%%%%%%%%%%%%%%%%%%%
%%%%%%%%%%%%%%%%%%%%%%%%%%%%%%%%%%%%%%%%%%%%%%%%%%%%%%%%%%

SAW amplification has also been studied theoretically by Yurchenko \etal~\cite{Yurchenko_AA5_57144_15}. The authors simulated a system consisting of multi-layer graphene at the surface of a CdS substrate, and calculated the changes in SAW amplitude induced by electric currents in graphene. In addition to Rayleigh modes, they also studied the effect in transverse acoustic waves polarized along the sample surface, known as  Blustein-Gulyaev (BG) modes. Although they predicted SAW amplification in both cases, the Rayleigh modes were more intensively amplified because their acoustic and piezoelectric fields are localized closer to the crystal surface than those of BG waves. In addition, the drift velocity threshold for the observation of SAW amplification was larger for the BG mode because it propagates faster than the Rayleigh one. Similar to Thalmeier \etal for the case of the graphene conductivity, cf. Fig.~\ref{fig_Thalmeier}, Yurchenko \etal also established two different regimes, depending on the relation between the SAW wavelength and the mean-free-path of the graphene charge carriers, $\ell$. For $2\pi\ell\ll\lSAW$, the threshold condition for graphene-induced SAW amplification was the same as for Schr\"odinger particles, $v_d>\vSAW$. The Dirac characteristics of the graphene carriers became relevant in the colissionless limit $2\pi\ell\gg\lSAW$, where the threshold condition went down to just $v_d>\vSAW/2$.

\subsection{Control of graphene electronic properties by SAW lattices}\label{sec_Graphenemodulation}

Surface acoustic waves can also be used to modulate the energy dispersion of graphene electronic excitations. As an example, Dietel \etal~\cite{Dietel_PRB86_115450_12} have proposed the use of periodic potentials in moving superlattices to modify the electronic dispersion of ballistic graphene close to the charge-neutrality point. In their theoretical study, the authors demonstrate that, as long as the velocity of the moving superlattice is much lower than the Fermi velocity of the charge carriers, it will induce an anisotropic renormalization of the Dirac cone similar to its static counterpart~\cite{Park_NP4_213_08, Park_PRL101_126804_08}. As SAWs move typically three orders of magnitude slower than the graphene charge carriers, the periodic fields accompanying a SAW are a promising candidate for the realization of such kind of superlattices. Here, the main challenge is the obtention of graphene with an electron mean-free-path much larger than the wavelength of the SAW, which lies typically in the micrometer range. This difficulty, however, could be overcome by using graphene encapsulated in h-BN, where excepcionally large mobilities have been reported~\cite{Banszerus_NL16_1387_16}.

%%%%%%%%%%%%%%
\begin{figure}
\centering
\includegraphics[width=0.6\linewidth]{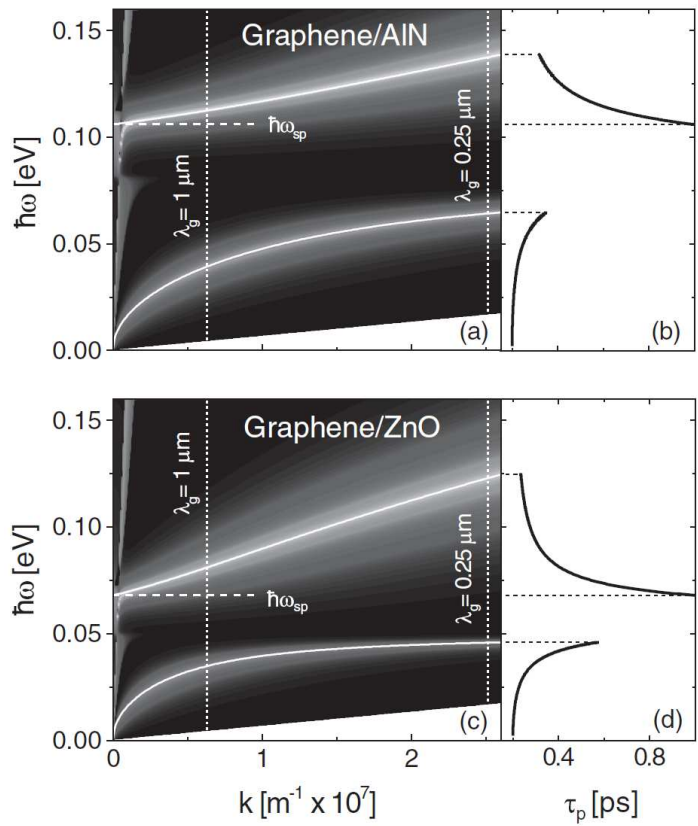}
\caption{Plasmon dispersion for graphene on (a) AlN, and on (c) ZnO. The contour plots correspond to the imaginary part of the light reflection coefficient. $\omega_{sp}$ denotes the frequency of the phonons located at the surface of the piezoelectric film (b) and (d) show the plasmon lifetime along the two hybridized plasmon branches. Reprinted figure with permission from~\cite{Schiefele_PRL111_237405_13}. Copyright (2018) by the American Physical Society.}
\label{fig_Schiefele}
\end{figure}
%%%%%%%%%%%%%%%%%%%%%%%%%%%%%%%%%%%%%%%%%%%%%%%%%%%%%%%%%%

Another application example of SAW-based superlattices is the coupling of laser light into graphene plasmons. The large mismatch between the wave vector of free-space radiation and that of the plasmons in graphene prevents their generation by optical means. However, such mismatch can be overcome by placing the graphene layer in a diffraction grating. The grating will scatter the incoming light into several diffraction orders with in-plane wave vectors, $k_{\parallel m}$, determined by:

\begin{equation}
k_{\parallel m}=(\omega/c)\sin\theta+m2\pi/\lambda_g.
\end{equation}

\noindent Here, $\omega$ and $c$ are the light frequency and velocity, respectively, $\theta$ is the angle of incidence with respect to the off-normal, $\lambda_g$ is the periodicity of the grating, and $m$ is an integer denoting the diffraction order. Plasmons will be excited in graphene when one of the possible values of $k_{\parallel m}$ matches the plasmon wave vector at the corresponding frequency. Schiefele \etal~\cite{Schiefele_PRL111_237405_13} have suggested that the periodic elastic deformations induced by SAWs can act as such diffraction grating with periodicity $\lambda_g=\lSAW$. In their proposed method, the frequency of the plasmons is tuned to the frequency of the light by controlling the graphene carrier density, while the SAW wavelength determines the range of plasmon wave vectors accessible via diffraction of the light beam by the SAW. The advantage of this method lies in the fact that tunable, SAW-induced gratings can extend over the full area covered by the graphene layer without the need of any complex near-field techniques~\cite{Chen_N487_77_12, Fei_N487_82_12} or patterning~\cite{Ju_NatNano6_630_2011, Yan_NatPhotonics7_394_2013}, thus avoiding plasmon damping due to scattering at the edges of the patterns. To demonstrate the feasibility of their approach, the authors have calculated the plasmon dispersion relation in graphene placed at the surface of two piezoelectric materials where SAWs can be efficiently generated, namely AlN and ZnO, cf. Figs.\ref{fig_Schiefele}(a) and \ref{fig_Schiefele}(c) respectively. Due to the coupling of the collective electronic excitations in graphene to polar surface phonons in the piezoelectric material, the plasmon dispersion relation splits into two hybridized branches. The corresponding plasmon lifetimes are displayed in Figs.~\ref{fig_Schiefele}(b) and \ref{fig_Schiefele}(d). At low wave vectors, the surface phonons dominate the character of the upper plasmon branch, thus allowing the excitation of relatively long-lived plasmons with almost 1~ps lifetimes. The lower plasmon branch, in contrast, is strongly influenced by electronic scattering, leading to shorter lifetimes. For normal incident light ($\theta=0$) and SAWs with $\lambda_g$ between 0.25~$\mu$m and 1~$\mu$m, $k_{\parallel 1}$ takes values between the two vertical dotted lines in Figs.~\ref{fig_Schiefele}(a) and \ref{fig_Schiefele}(c), well within the range of wave vectors where plasmons can be efficiently excited. SAWs with such short wavelengths can nowadays be generated by IDT fingers patterned using either electron-beam or nano-imprint lithography~\cite{PVS250}.

%%%%%%%%%%%%%%%%%%%%%%%%%%%%%%%%%%%%%%%%%%%%%%%%%%
\section{SAW-induced acousto-electric currents}\label{sec_AEcurrents}

\subsection{CVD graphene transferred to LiNbO$_3$}\label{sec_transferred}

\begin{figure}
\centering
\includegraphics[width=0.6\linewidth]{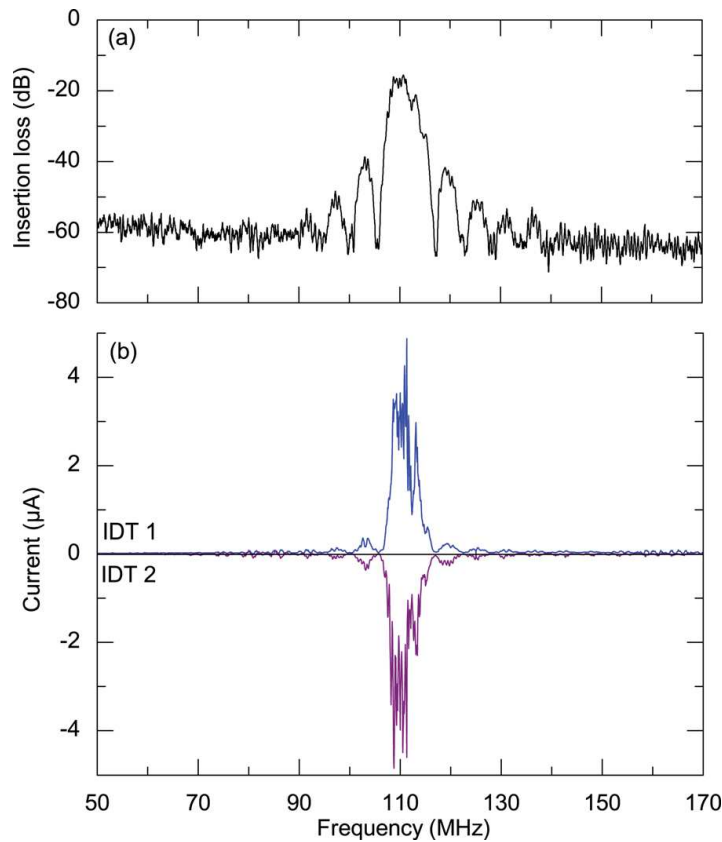}
\caption{(a) Transmission coefficient, $\st$, between IDT1 and IDT2 patterned at either side of a graphene film. (b) Acousto-electric current measured between two contacts of the graphene film for SAWs generated by IDT1 (upper panel) and IDT2 (lower panel), and therefore travelling along opposite directions. Reprinted from~\cite{Miseikis_APL100_133105_12} with the permission of AIP Publishing.}
\label{fig_Miseikis2012}
\end{figure}

As mentioned in Sec.~\ref{sec_SAWmodification}, the interaction between SAWs and a two-dimensional electron gas typically induces a transfer of kinetic energy from the acoustic wave into the charge carriers that results in the attenuation of the SAW. From the point of view of the electron gas, this energy transfer shows up as an electric current due to the acousto-electric (AE) effect~\cite{Parmenter_PR89_990_53, Weinreich_PR104_321_56, Ingebrigtsen_JAP41_454_70}. The first experimental demonstration of AE currents in graphene was reported by Miseikis \etal~\cite{Miseikis_APL100_133105_12}. The authors patterned two IDTs on a $128^\circ$ Y-rotated single crystal LiNbO$_3$ for the generation and detection of SAWs, and deposited a CVD-grown graphene layer between them. Figure~\ref{fig_Miseikis2012}(a) displays the insertion loss of the SAW delay line, while Fig.~\ref{fig_Miseikis2012}(b) shows the  frequency dependence of the AE current measured along the graphene film. It follows closely the response of the SAW delay line and reproduces not only the main peak of the insertion loss, but also the secondary peaks at both sides. Moreover, the AE current changed sign when the SAW was excited using the opposite IDT, thus confirming the dependence of the carrier propagation direction on that of the SAW. 

%%%%%%%%%%%%%%
%%%%%%%%%%%%%%%%%%%%%%%%%%%%%%%%%%%%%%%%%%%%%%%%%%%%%%%%%%

\begin{figure}
\centering
\includegraphics[width=0.6\linewidth]{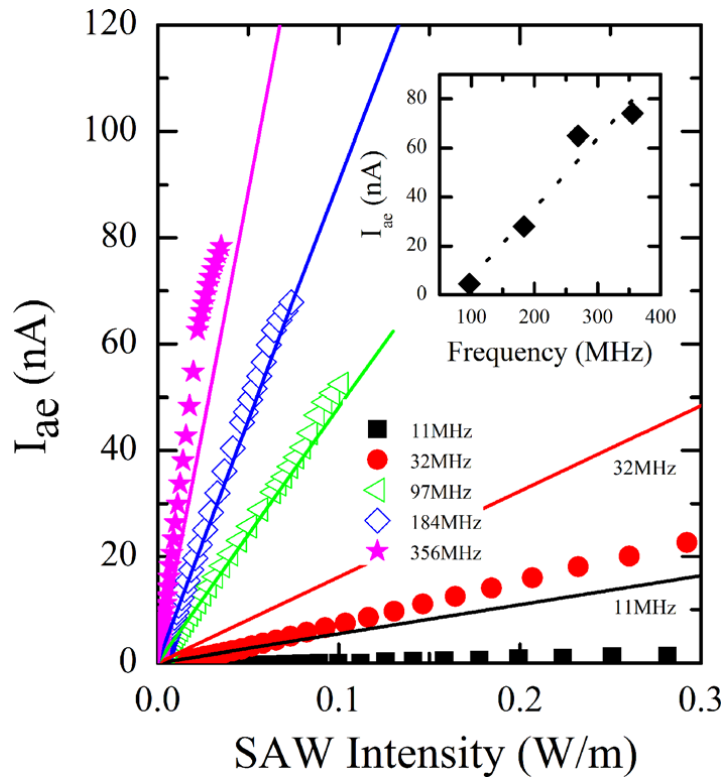}
\caption{(a) Measured (symbols) and calculated (lines) acousto-electric current, $\Iae$, in graphene on LiNbO$_3$ as a function of SAW intensity and frequency. The inset shows the amplitude of $\Iae$ as a function of the SAW frequency, measured at a SAW intensity of 0.03 W/m. The dotted line is a fit to the data. Reprinted from~\cite{Bandhu_APL103_133101_13} with the permission of AIP Publishing.}
\label{fig_Bandhu2013}
\end{figure}

Miseikis \etal also observed that the amplitude of the AE current varied linearly with the power of the SAW. This observation was confirmed shortly afterwards by Bandhu~\etal~\cite{Bandhu_APL103_133101_13}, who studied AE currents in CVD graphene transferred to LiNbO$_3$ under several SAW frequencies. Figure \ref{fig_Bandhu2013} displays the amplitude of the AE current, $\Iae$, as a function of the SAW intensity. For all measured SAW frequencies, $\Iae$ increased linearly with SAW power. These results agree well with the classical relaxation model employed to describe AE currents in semiconductor heterostructures~\cite{Esslinger_SSC84_939_92, Shilton_PRB51_14770_95, Shilton_JoPCM7_7675_95, Rotter_APL73_2128_98}. According to this model, $\Iae$ can be expressed, under short-circuit conditions and in absence of an external magnetic field, as:

\begin{equation}\label{eq_AEcurrent}
\Iae=-\mu\frac{\PSAW}{\vSAW}\alpha.
\end{equation}

\noindent Here, $\mu$ is the carrier mobility, $\PSAW$ and $\vSAW$ are the SAW power and velocity, respectively, and $\alpha$ is the SAW attenuation ratio discussed in Eq.~\ref{eq_SAWobservables}, which can be rewritten as:

\begin{equation}\label{eq_Attenuation}
\alpha=q\frac{\keff}{2}\frac{\sigma/\sM}{1+(\sigma/\sM)^2}.
\end{equation}

\noindent From these relations, one can deduce that the amplitude of the AE current depends linearly not only on the SAW power, but also on the frequency of the SAW via the wave vector, $q=\fSAW/\vSAW$. This dependence has also been confirmed experimentally, cf. inset of Fig.~\ref{fig_Bandhu2013}, which displays the value of $\Iae$ recorded under the same SAW intensity, but using different SAW frequencies. Similar results have been reported for graphene patterned into nano-ribbons~\cite{Poole_SR7_1767_17}. In that case, a comparison of the AE currents measured at nano-ribbons with different widths showed that, in general, $\Iae$ increased as the ribbon width decreased, thus indicating an enhancement of the carrier mobility for the narrower ribbons~\cite{Iqbal_ACSApplMat&Interf6_4207_2014, Lee_MicroelEng163_55-59_2016}.

%%%%%%%%%%%%%%
%%%%%%%%%%%%%%%%%%%%%%%%%%%%%%%%%%%%%%%%%%%%%%%%%%%%%%%%%%

Contrary to the SAW power and frequency, the classical relaxation model predicts a non-monotonic dependence of the AE current on the conductivity of the electron gas. This behavior contrasts to that of conventional DC currents, which are always proportional to the conductivity. According to Eqs.~\ref{eq_AEcurrent} and~\ref{eq_Attenuation}, the AE current is initially proportional to $\sigma$. It reaches a maximum, however, when $\sigma$ equals $\sM$, and decreases when $\sigma$ moves towards values larger than $\sM$. As mentioned previously, $\sM$ is a characteristic conductivity that depends on the materials surrounding the 2DEG. It is defined as $\sM=\vSAW(\varepsilon_1+\varepsilon_2)$, where $\varepsilon_1$ and $\varepsilon_2$ are the dielectric constants of the materials directly below and above the 2DEG. Therefore, the selection of the materials that will be in contact with the graphene film is an important issue in the design of acousto-electric devices, as it will have a strong impact on the value of the electric conductivity that maximizes the coupling between the SAW and the electron gas.

\begin{figure}
\centering
\includegraphics[width=0.6\linewidth]{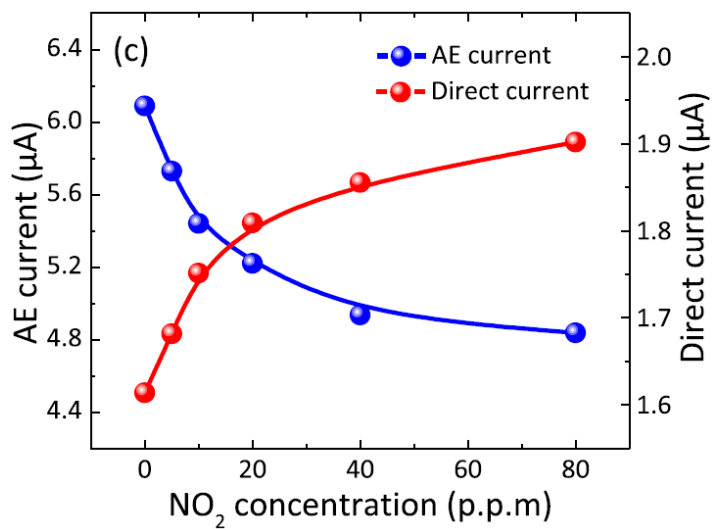}
\caption{Amplitude of the AE current (blue points) and conventional direct current (red points) measured under exposition of graphene to several concentrations of a NO$_2$ gas dopant. Reprinted from~\cite{Zheng_APL109_183110_16} with the permission of AIP Publishing.}
\label{fig_Zheng}
\end{figure}

To demonstrate the special dependence of the AE current on the electric conductivity, the graphene carrier density has been modulated by several means, e.g. photoexcitation of electron-hole pairs~\cite{Poole_APL106_133107_15, Poole_JPD51_154001_2018} and chemical doping~\cite{Zheng_APL109_183110_16}. Figure~\ref{fig_Zheng} shows an example of the last case, where a p-doped graphene layer was exposed to different concentrations of nitric acid (NO$_2$) vapor. As mentioned in Sec.~\ref{sec_SAWs}, the NO$_2$ molecules adsorbed by the graphene lead to the enhancement of its p-doping character and therefore to the increase of its conductivity. As a result, the amplitude of the conventional current induced in graphene increased with NO$_2$ concentration (red points). The AE current, on the contrary, decreased because $\sigma$, which was already larger than $\sM$ before the exposition to the vapor, moved further away from $\sM$ as the gas concentration increased (blue points). When NO$_2$ was replaced by a donor of electrons like NH$_3$, the partial compensation of the graphene p-doping lead to the decrease of $\sigma$. As a consequence, the amplitude of the conventional current reduced, while the AE current increased with the concentration of NH$_3$, in agreement with $\sigma$ approaching $\sM$ (not shown here). Remarkably, for low gas concentrations, the relative response changes of the AE current outperformed that of the common drift current~\cite{Zheng_APL109_183110_16}, thus demonstrating the potential application of graphene-based acousto-electric devices for chemical detection.

%%%%%%%%%%%%%%
%%%%%%%%%%%%%%%%%%%%%%%%%%%%%%%%%%%%%%%%%%%%%%%%%%%%%%%%%%

In contrast to photoexcitation and chemical doping, which only induced small variations in the graphene conductivity, the modulation of $\Iae$ in a broad range of carrier densities has been demonstrated by means of a field-effect technique. To this end, a top gate consisting of a high-capacitance ion-gel dielectric~\cite{Bandhu2016} or an electrolytic solution~\cite{Tang_JoAP121_124505_17} was deposited on the graphene/LiNbO$_3$ system. Okuda \etal~\cite{Okuda_APE9_45104_16, Okuda2018}, on the contrary, worked with graphene on LiTaO$_3$ instead of LiNbO$_3$ because the shear-horizontal SAWs propagating in LiTaO$_3$ minimize the leaking of the SAW energy into the ionic liquid. Recently, modulation of the AE current using a bottom gate has also been reported. In this case, a strong piezoelectric layer like AlN~\cite{Liang2017} or LiNbO$_3$~\cite{Liang_JPhysD51_204001_2018} placed above the gate electrode and below the graphene layer acted as both the piezoelectric material for the acoustic device and the dielectric film of the field-effect transistor.

%%%%%%%%%%%%%%
\begin{figure}
\centering
\includegraphics[width=0.6\linewidth]{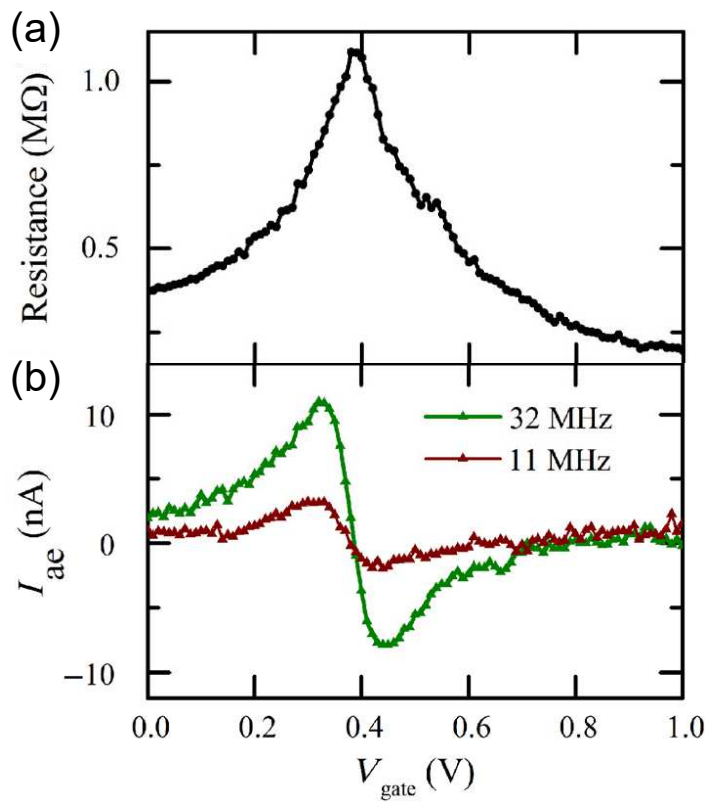}
\caption{(a) Two-contact DC-resistance of a graphene layer on LiNbO$_3$ as a function of the voltage bias, $\Vg$, applied to an ion-gel dielectric acting as top gate. (b) Acousto-electric current, $\Iae$, as a function of $\Vg$ for two SAW frequencies. Reprinted with modifications from \cite{Bandhu2016} under the terms of the Creative Commons Attribution 4.0 International License (http://creativecommons.org/licenses/by/4.0/).}
\label{fig_Bandhu2016}
\end{figure}
%%%%%%%%%%%%%%%%%%%%%%%%%%%%%%%%%%%%%%%%%%%%%%%%%%%%%%%%%%

Figure~\ref{fig_Bandhu2016}(a) displays the two-contact DC-resistance of a graphene layer on LiNbO$_3$ as a function of the bias voltage applied to the top gate, $\Vg$~\cite{Bandhu2016}. The maximum value, observed at $V_0\approx0.4$~V, indicates the gate voltage for which the chemical potential reaches the charge-neutrality point. Figure~\ref{fig_Bandhu2016}(b) shows the dependence of the AE current on $\Vg$ for two different SAW frequencies, 11~MHz (brown triangles) and 32~MHz (green triangles). In absence of gate voltage, the AE current is positive due to the natural p-doping of the graphene layer. The application of a positive $\Vg$ moves the chemical potential towards the charge-neutrality point, thus reducing the graphene conductivity. As $\sigma$ was larger than $\sM$ in the absence of the gate voltage, $\Iae$ increases first to a maximum at $\Vg\approx0.33$~V, where the graphene conductivity equals $\sM$, and then falls until it reaches zero at the charge-neutrality point. Gate voltages above $V_0$ increase the conductivity again, but changing the graphene doping from holes into electrons. Therefore, the same features are observed for $\Vg>V_0$ as for $\Vg<V_0$, but now the sign of $\Iae$ is negative because both electrons and holes propagate in the same direction as the SAW. This is also a characteristic feature of the acousto-electric current, because conventional DC currents do not reverse sign under change of the carrier polarity. 

Remarkably, $\Iae$ vanishes at $V_0$ in spite of the fact that the conductivity measured at the charge-neutrality point is, in general, well above zero. The conductivity of large-area CVD graphene is determined by many competing mechanisms~\cite{Heo2011}. In particular, non-homogeneous charged impurities can lead to a non-zero DC-conductivity at the charge-neutrality point due to the formation of electron-hole puddles~\cite{Adam_PNAS104_18392_07, Li_JoAP114_233703_13}. Recently, Tang \etal have reported suppression of $\Iae$ at $V_0$ even though the graphene conductivity at the charge-neutrality point exceeded $\sM$~\cite{Tang_JoAP121_124505_17}. They attributed this behavior to the opposite contribution to $\Iae$ of puddles with different polarity.

%%%%%%%%%%%%%%
\begin{figure}
\centering
\includegraphics[width=0.6\linewidth]{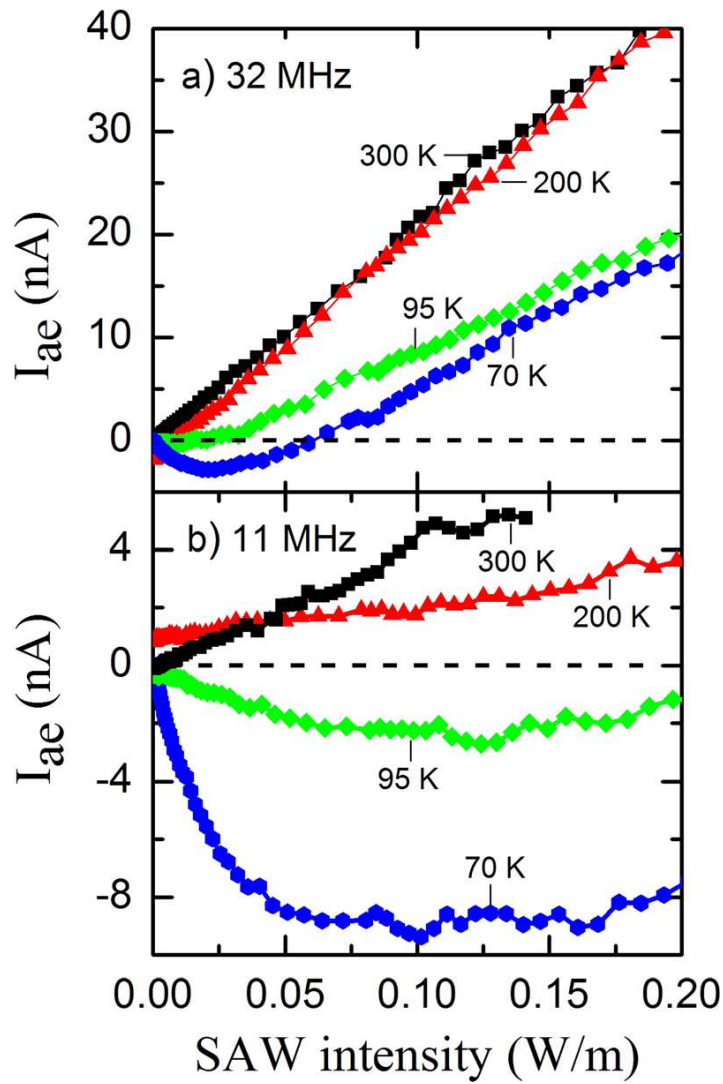}
\caption{Acousto-electric current measured as a function of SAW intensity and temperature for SAW frequencies of (a) 32~MHz and (b) 11~MHz. Reprinted from~\cite{Bandhu_APL105_263106_14} with the permission of AIP Publishing.}
\label{fig_Bandhu2014}
\end{figure}
%%%%%%%%%%%%%%%%%%%%%%%%%%%%%%%%%%%%%%%%%%%%%%%%%%%%%%%%%%

The AE current is also sensitive to additional mechanisms affecting the electronic properties of graphene. For example, it has been reported that the AE current decreased in CVD graphene when cooling the sample from room temperature to that of liquid nitrogen~\cite{Bandhu_APL105_263106_14}. The measurement was well-described by a thermally activated electronic mobility following an Ahrrenius law. It is well known that such a dependence is caused by potential barriers related to defects in graphene like grain boundaries, tears, wrinkles, as well as the previously mentioned charged impurities~\cite{Zhang2009, Li2011, Song2012, Kumari2014}. At low temperatures and carrier densities, the conductivity of CVD graphene depends strongly on the percolation of thermally activated carriers across such inhomogeneous potential fluctuations, thus affecting the behavior of the AE current. Figure~\ref{fig_Bandhu2014} displays $\Iae$ as a function of SAW intensity for (a) $\fSAW=32$~MHz, and (b) $\fSAW=11$~MHz, measured at several temperatures. At room temperature, $\Iae$ is positive and increases linearly with SAW power, as predicted by Eq.~\ref{eq_AEcurrent}. At temperatures below 100~K, however, $\Iae$ does not show a linear dependence on SAW power anymore. Moreover, it becomes negative within certain range of SAW intensities, thus indicating that, in overall, the SAW is now transporting more electrons than holes. These results have been attributed to the fact that, as previously discussed in Eq.~\ref{eq_SAWobservables}, the attenuation of the SAW depends actually on the electronic conductivity at the corresponding SAW frequency and wave vector. Therefore, the use of several SAW frequencies is equivalent to the probing of the conductivity over several length scales, determined by the different SAW wavelengths. While high frequency SAWs probe the graphene conductivity over length scales typically smaller than the average crystal grain size, low frequency SAWs are more sensitive to the potential fluctuations present in the graphene layer because of the larger length scales over which they prove the conductivity.

\subsection{Epitaxial graphene on SiC}\label{sec_epitaxial}

The functionalities discussed in the previous sections have been demonstrated in graphene layers mechanically transferred to the surface of piezoelectric materials. For the future application of such functionalities in graphene-based acousto-electric devices, however, it is desirable to use material combinations that allow for large scale fabrication at relatively low costs. From this point of view, epitaxial graphene (EG) on SiC is a promising candidate because it makes possible the preparation and patterning of large-area graphene layers directly on an insulating substrate. Moreover, the large acoustic velocity of SiC (cf. Table~\ref{tab_substrates}) allows the propagation of SAWs with larger frequencies than in e.g. LiNbO$_3$ using the same SAW wavelength. The main disadvantage of SiC lies in the fact that it is a weak piezoelectric: its electro-mechanical coupling coefficient is almost 500 times weaker than that of LiNbO$_3$, cf. Table~\ref{tab_substrates}. Therefore, the realization of SAW-based functionalities in EG on SiC requires the addition of piezoelectric layers for both the generation of SAWs and the efficient coupling of the graphene charge carriers to the SAW piezoelectric field.

%%%%%%%%%%%%%%
\begin{figure}
\centering
\includegraphics[width=0.6\linewidth]{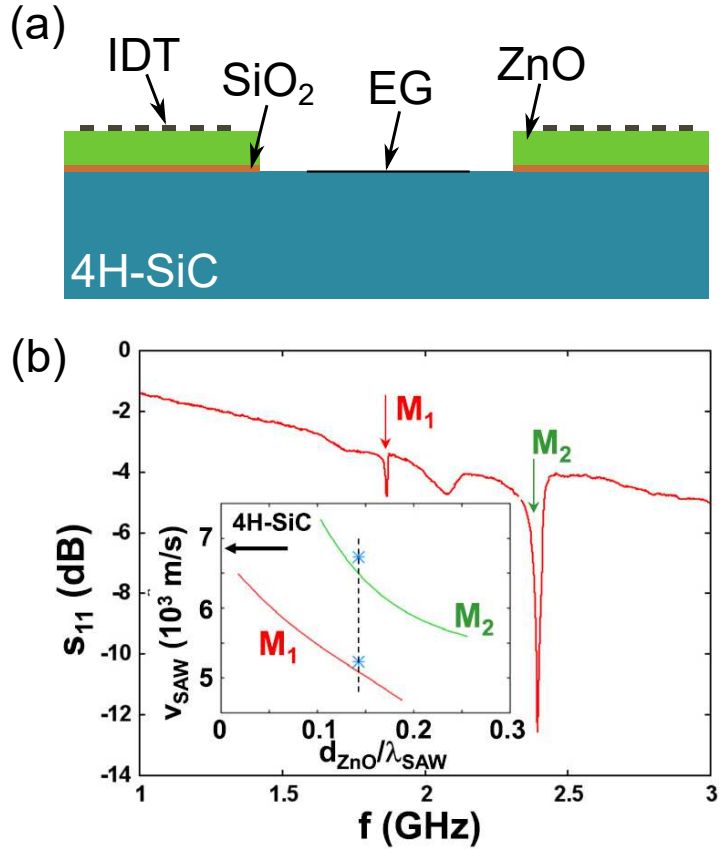}
\caption{(a) Schematic cross-sectional view of a device for SAW-induced generation of acousto-electric currents in EG on SiC. The IDTs are deposited on piezoelectric islands consisting of a 50~nm SiO$_2$ buffer layer and a 350~nm-thick ZnO film.  (b) frequency-resolved rf-power reflection coefficient, $\sr1$, of the delay line. The IDTs generate two Rayleigh modes, $M_1$ and $M_2$ (indicated by the vertical arrows). The inset displays the measured (symbols) and the numerically calculated (lines) velocities of the two modes as a function of the ZnO thickness, $d_\mathrm{ZnO}$. The horizontal arrow indicates the SAW velocity in 4H-SiC. Reprinted from~\cite{Santos_APL102_221907_13} with the permission of AIP Publishing.}
\label{fig_Santos2013a}
\end{figure}
%%%%%%%%%%%%%%%%%%%%%%%%%%%%%%%%%%%%%%%%%%%%%%%%%%%%%%%%%%

The generation of AE currents in EG on SiC was first reported by Santos \etal~\cite{Santos_APL102_221907_13}. Figure~\ref{fig_Santos2013a}(a) displays a side view schematic structure of the device used in the experiment. The authors first structured the graphene layer in the form of 10~$\mu$m-wide Hall bars with Ti/Au contacts. Next, they sputtered a 50~nm-thick SiO$_2$ buffer layer and 350~nm of a piezoelectric ZnO film on the sample surface. As the sputtering process damages graphene, they used a shadow mask to deposit the SiO$_2$/ZnO bilayer in areas away from the graphene structures. Finally, they patterned IDTs for the generation of SAWs with $\lSAW=2.8~\mu$m on top of these piezoelectric islands. Figure~\ref{fig_Santos2013a}(b) displays the frequency-resolved $\sr1$ coefficient of the rf-signal applied to the IDTs. In addition to the fundamental SAW mode ($M_1$) with 1.87~GHz frequency, the IDTs generate a faster overtone ($M_2$) at 2.40~GHz. This is due to the fact that the acoustic mismatch between ZnO and SiC (as can be seen in Table~\ref{tab_substrates}, SAWs propagate more than two times faster in SiC than in ZnO) acts as a two-dimensional wave guide, thus strongly confining the SAW at the top layer and giving rise to additional, faster SAW overtones~\cite{Didenko_IToUFaFC47_179_00}. The number of such overtones depends on the thickness of the ZnO layer, $d_\mathrm{ZnO}$. In absence of ZnO, only $M_1$ is a surface solution of the elastic equations. In contrast, for ZnO layers approaching $d_\mathrm{ZnO}\approx\lSAW$, at least four SAW overtones can be generated~\cite{PVS250}. The inset of Fig.~\ref{fig_Santos2013a}(b) compares the measured (symbols) and the numerically calculated (lines) phase velocities of the $M_1$ and $M_2$ modes as a function of the ZnO layer thickness, indicating a reasonable agreement between theory and experiment.

%%%%%%%%%%%%%%
\begin{figure}
\centering
\includegraphics[width=0.6\linewidth]{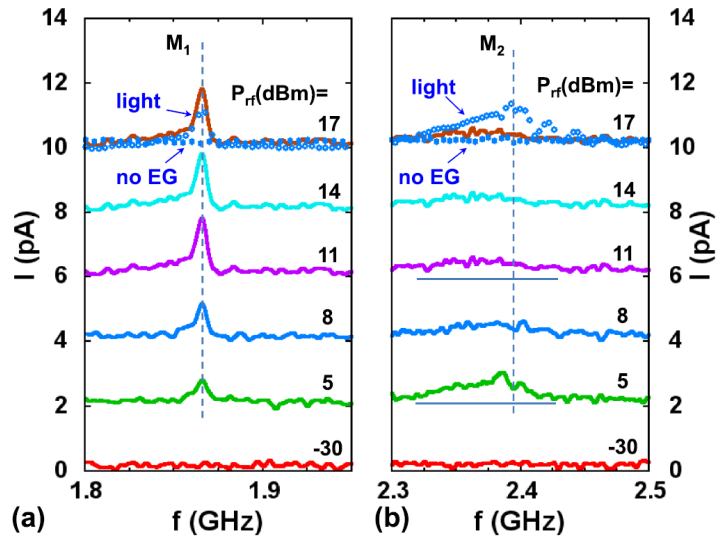}
\caption{Spectral dependence of the current (I) measured between two Hall bar contacts as a function of the nominal rf-power ($\Prf$) applied to the IDT for the modes (a) $M_1$ and (b) $M_2$. The symbols denoted by "light" (open circles) and "no EG" (dots) display the curent measured under illumination and on a control sample where the epitaxial graphene (EG) was removed during processing, respectively. The curves are displayed vertically for clarity. Reprinted from~\cite{Santos_APL102_221907_13} with the permission of AIP Publishing.}
\label{fig_Santos2013b}
\end{figure}
%%%%%%%%%%%%%%%%%%%%%%%%%%%%%%%%%%%%%%%%%%%%%%%%%%%%%%%%%%

Figure~\ref{fig_Santos2013b} displays the dependence of the AE current measured in the graphene Hall bar on the rf-frequency applied to the IDT for both SAW modes described in Fig.~\ref{fig_Santos2013a}(b). As expected, the current induced by mode $M_1$ increases proportionally to the nominal power $\Prf$ applied to the IDT. In contrast, the acousto-electric current associated to mode $M_2$ has a more complicated dependence on the rf-power. It initially increases, reaches a maximum, and then reduces for larger $\Prf$. Moreover, the electric current induced by this mode is much more sensitive to illumination than in the case of mode $M_1$ (cf. the curves marked ``light"). To confirm that the conduction of charge carriers takes place through the graphene rather than through the underlying substrate, a control measurement was carried out on a sample, where the graphene layer was removed prior to the deposition of metal contacts (curve "no EG" for $\Prf=17$~dBm). The current completely vanished in this case.

\begin{figure}
\centering
\includegraphics[width=0.6\linewidth]{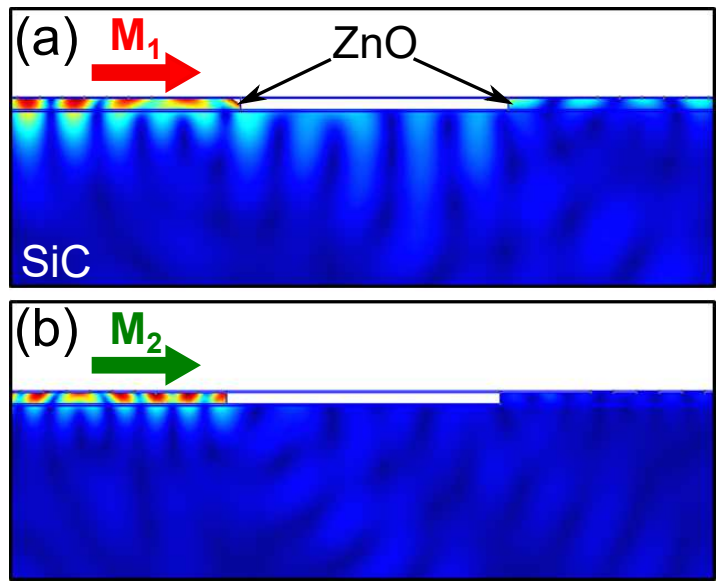}
\caption{Numerically calculated mechanical displacement of the SAW modes (a) $M_1$ and (b) $M_2$ for the ZnO/SiO$_2$/SiC multilayer of Fig.~\ref{fig_Santos2013a}. The color scale indicates the absolute amplitude of $\vec{u}$, normalized to its value at the output of the IDT that generates the SAW (left side edge of the images).}
\label{fig_sim1}
\end{figure}

The AE current measured for the $M_2$ mode is, in general, weaker than for the $M_1$ mode. This contrasts with the results for the $\sr1$ coefficient, where the largest dip, and therefore the strongest SAW generation efficiency, corresponds to the $M_2$ mode, c.f. Fig.~\ref{fig_Santos2013a}(b). The reason for this discrepancy is the fact that the SAWs are generated at the ZnO islands, but the graphene structures are located at the area without ZnO, where the propagation behavior is different for each mode. To illustrate this, we have numerically simulated the spatial distribution of the mechanical displacement for both SAW modes using a finite element approach~\cite{getdp,gmsh_general}. We assumed a two-dimensional model for the acoustic delay line corresponding to a $x$-$z$ cross-section of the layered structure, and used the same ZnO thickness and SAW wavelength as in the real sample. To make the problem numerically tractable, we reduced the number of IDT fingers to 20 pairs and the region without ZnO to just $3\lSAW$. Figure~\ref{fig_sim1}(a) displays the absolute amplitude of the mechanical displacement for the $M_1$ mode, generated by an IDT deposited on the ZnO island shown at the left of the image. Figure~\ref{fig_sim1}(b) shows the results of the simulation for the $M_2$ mode. Both modes are confined at the ZnO layer, but the $M_1$ mode penetrates deeper into the SiC substrate than the $M_2$ mode. Although the transition out of the ZnO island causes scattering and attenuation in both cases, the $M_1$ mode also propagates along the surface of the non-coated SiC region, thus reaching the ZnO island at the right side of the image. In contrast to the $M_1$ mode, the efficient propagation of the $M_2$ mode is strongly bound to the presence of the ZnO film. When the $M_2$ mode leaves the ZnO island, only a small fraction of the SAW energy couples into the non-coated SiC region, where it becomes a pseudo-SAW that penetrates into the bulk. This explains why the AE currents measured in Fig.~\ref{fig_Santos2013b} are weaker for the $M_2$ mode than for the $M_1$ one. 

%%%%%%%%%%%%%%%%%%
%%%%%%%%%%%%%%%%%%%%%%%%%%%%%%%%%%%%%%%%%%%%%%%%%%%%%%%%%%

Figure~\ref{fig_sim1} also suggests that the efficiency of the AE interaction will significantly improve if the ZnO layer is sputtered not only below the IDTs, but also on top of the EG. In addition to avoid the SAW losses induced at the transition out of the ZnO islands, the strong piezoelectric fields generated by the ZnO layer are expected to induce more intense AE currents in graphene than the currents generated by the weak piezoelectric field of the SiC. This approach, however, requires the incorporation of a protective interlayer on the EG because the sputtering process used for the deposition of the ZnO film damages graphene. Hern\'andez-M\'inguez \etal have reported AE currents in graphene covered by ZnO, where they used a 100~nm-thick hydrogen-silsesquioxane (HSQ) film as protective interlayer~\cite{Hernandez-Minguez_APL108_193502_16}. To this purpose, they spin-coated the HSQ on the graphene structures and then baked the sample to solidify the HSQ. Raman characterization after fabrication confirmed that HSQ preserved the EG during the sputtering of the ZnO film. However, the graphene carrier mobility, measured in field-effect transistors using the layers on the EG as the dielectric material for the top metal gate, was just about 100 cm$^2$/Vs. This mobility is one order of magnitude lower than the one typically measured in non-covered EG on SiC~\cite{Jobst_SSC151_1061_11}. Nevertheless, the strong SAW piezoelectric fields generated by the ZnO layer induced AE currents in graphene reaching almost 1~nA for the largest rf-power applied to the IDT. 

%%%%%%%%%%%%%%
\begin{figure}
\centering
\includegraphics[width=0.6\linewidth]{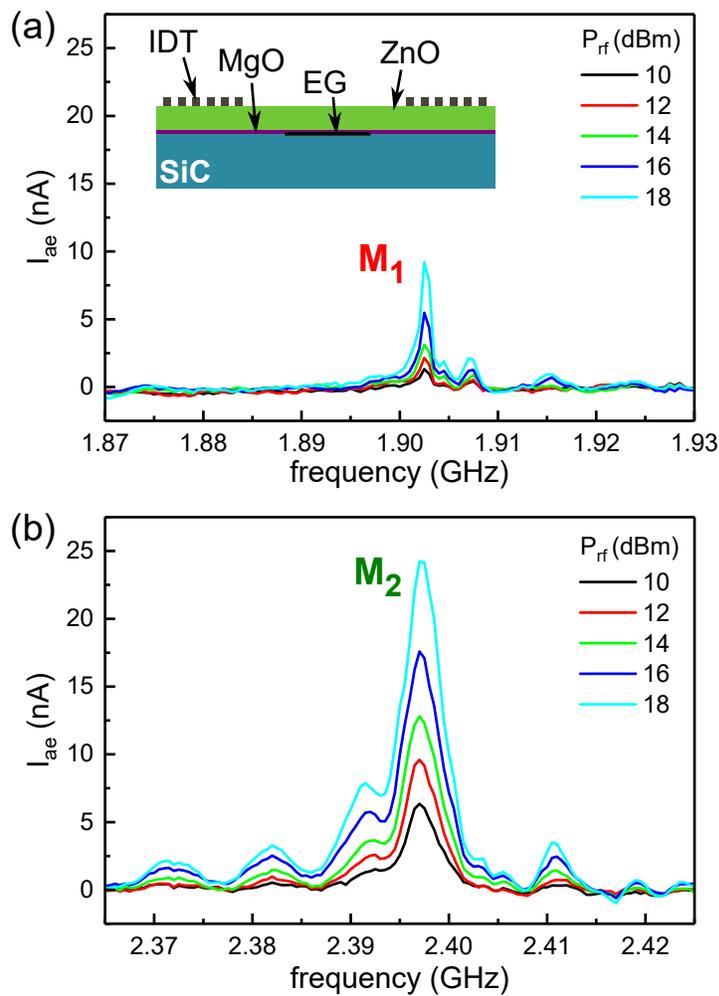}
\caption{Amplitude of the acousto-electric current, $\Iae$, measured in a graphene stripe coated with MgO/ZnO as a function of the rf-frequency applied to the IDT. The experiment was performed at frequencies around the SAW modes (a) $f_{M1}=1.9$~GHz and (b) $f_{M2}=2.4$~GHz, and for several nominal input rf-powers, $\Prf$. Figure adapted from Liou \etal~\cite{Liou_JoPDAP50_464008_17}.}
\label{fig_Liou}
\end{figure}
%%%%%%%%%%%%%%%%%%%%%%%%%%%%%%%%%%%%%%%%%%%%%%%%%%%%%%%%%%

Recently, Liou \etal~\cite{Liou_JoPDAP50_464008_17} have demonstrated a significant improvement in the generation of SAW-induced electric currents in EG on SiC by replacing the relatively thick HSQ interlayer with a thin MgO film (see inset of Fig.~\ref{fig_Liou}(a)). MgO has been identified as a promising insulator in graphene-based devices, e.g. as efficient tunnel barrier between graphene and ferromagnetic contacts for injection of spin-polarized electrons~\cite{PhysRevLett.105.167202}. In their experiment, the authors deposited a 15~nm-thick MgO layer on the graphene structures by sublimating pure Mg at 350 $^\circ$C and providing molecular oxygen in an ultra-high vacuum chamber. After sputtering the ZnO film, they measured a field-effect mobility of 2900 cm$^2$/Vs at room temperature, which increased up to 3900 cm$^2$/Vs when measured at cryogenic temperature (80 K). These findings indicate that the MgO-coating effectively protects graphene and preserves its electronic properties. Figure~\ref{fig_Liou} displays the AE currents measured for the two Rayleigh modes generated in the ZnO/MgO/SiC multilayer. For the low frequency mode ($M_1$), the amplitude of $\Iae$ reached 10~nA at the highest $\Prf$ used, see Fig.~\ref{fig_Liou}(a). This is one order of magnitude larger than the values achieved, under similar rf-frequencies and powers, using HSQ as protective layer. As the authors used the same IDT design and the same nominal thickness of the ZnO film in both HSQ- and MgO-coated samples, they attributed the different amplitude of the acousto-electric current to the better electronic mobility of EG coated by MgO with respect to graphene coated by HSQ. An even better performance was observed for the SAW mode $M_2$, cf. Fig.~\ref{fig_Liou}(b), where $\Iae$ reached 25~nA for the largest $\Prf$ used. This agrees with the behavior expected from the $\sr1$ measurement of Fig.~\ref{fig_Santos2013a}(b), thus confirming the better generation efficiency of the $M_2$ mode with respect to the $M_1$ mode for the thickness of the ZnO film used in the acoustic device. 

%%%%%%%%%%%%%%
\begin{figure}
\centering
\includegraphics[width=0.6\linewidth]{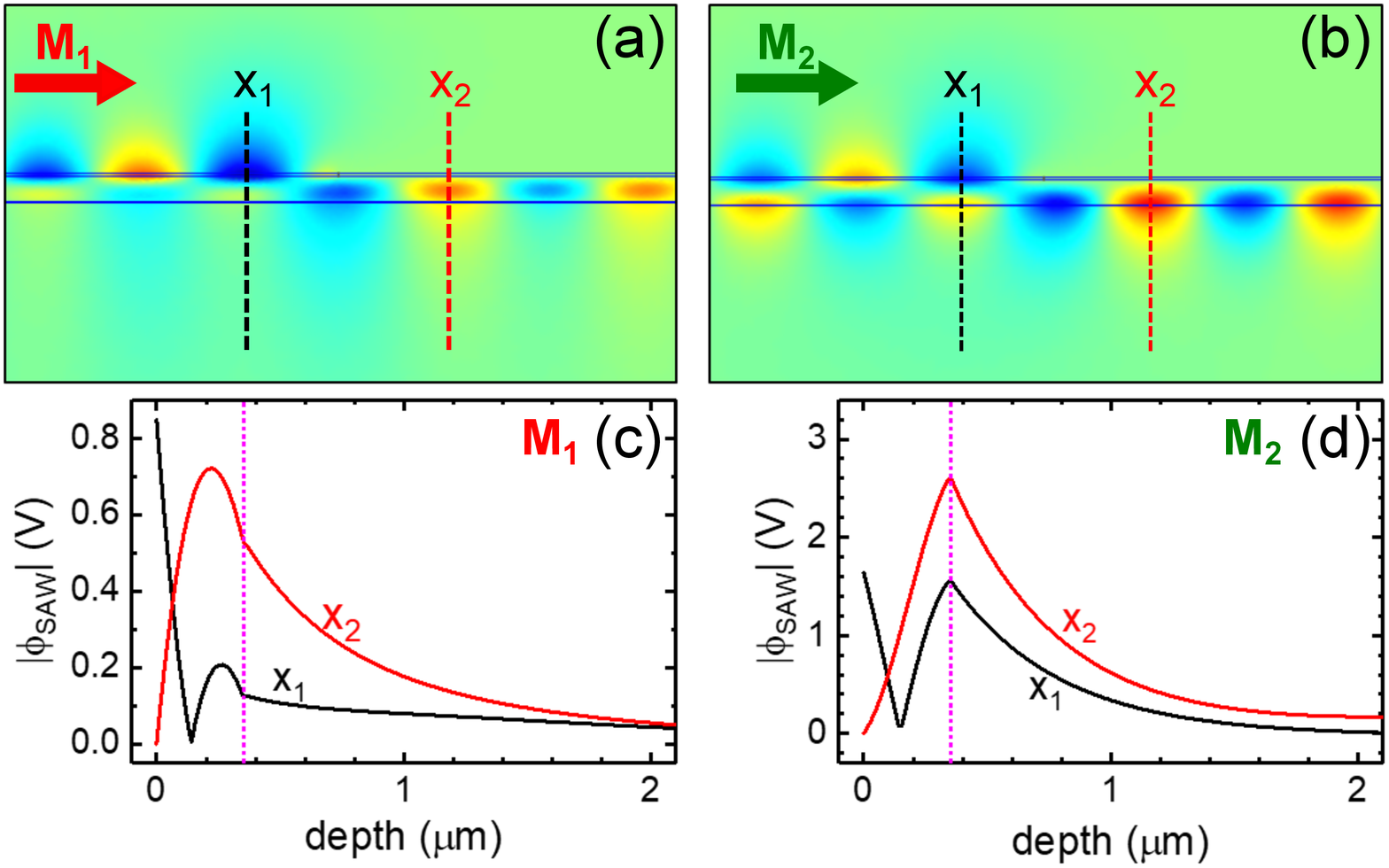}
\caption{(a) Numerically calculated amplitude of the piezoelectric potential, $\VSAW$, for the SAW mode M1 as it penetrates into a region with a metal on top of the ZnO film. (b) Same as (a) for the M2 mode. (c-d) Vertical profile of $|\VSAW|$ at positions $x_1$ (without top metal) and $x_2$ (with top metal) for both modes.}
\label{fig_sim2}
\end{figure}
%%%%%%%%%%%%%%%%%%%%%%%%%%%%%%%%%%%%%%%%%%%%%%%%%%%%%%%%%%

As mentioned previously, the ZnO can also act as the dielectric material in a top metal gate for the control of the graphene carrier density. This scheme would allow the modulation of the AE current in EG on SiC in a similar way as has already been demonstrated in CVD graphene transferred to a piezoelectric substrate. It must be taken into account, however, that the presence of the metal layer changes the boundary conditions of the coupled acoustic and electric equations, and that this modification can have important consequences in the spatial distribution of the SAW piezoelectric fields. To study this in detail, we have calculated, using the same numerical method as in Fig.~\ref{fig_sim1}, the amplitude of the SAW piezoelectric potential, $\VSAW$, for an acoustic wave traveling along a ZnO/SiC bilayer which was covered partially by an Al metal film. In our calculations, we do not take into account the contribution of the graphene carrier density to the total potential, as the SAW experiments directly measure the response of the carrier density to the perturbing externally applied scalar potential~\cite{Simon_PRB54_13878_96}. Figures~\ref{fig_sim2}(a) and \ref{fig_sim2}(b) display the results of the simulation for modes $M_1$ and $M_2$, respectively, where the top metal film is placed at the right side of the images. In both cases, the piezoelectric field at the surface of the sample is screened by the free charges of the metal. This is accompanied by an enhancement of $\VSAW$ around the interface between the ZnO layer and the SiC substrate. This effect is seen more clearly at Figs.~\ref{fig_sim2}(c) and \ref{fig_sim2}(d), which display $|\VSAW|$ as a function of sample depth at the positions along the SAW propagation path marked as $x_1$ (without top metal) and $x_2$ (with top metal) in panels (a) and (b). The vertical dotted line indicates the depth of the ZnO/SiC interface, which is also the depth position of the graphene layer. As the amplitude of the AE current depends on the response of the carrier density to the value of $\VSAW$ at the depth of the two-dimensional gas~\cite{Simon_PRB54_13878_96}, the results of Fig.~\ref{fig_sim2} indicate that the presence of the top metal gate will further enhance the intensity of the SAW-induced acousto-electric current in ZnO-coated epitaxial graphene on SiC.

%%%%%%%%%%%%%%%%%%%%%%%%%%%%%%%%%%%%%%%%%%%%%%%%%%
\section{Conclusions and Outlook}\label{sec_Concl}

In this topical review, we have discussed the state of the art of the interaction between graphene electronic excitations and surface acoustic waves. After an introduction to the special electronic properties of graphene and the main characteristics of SAWs, we have first discussed the effect of the graphene electron gas on the SAW propagation velocity and attenuation. Due to the Dirac nature of the graphene electronic dispersion, lossless propagation of the SAW along graphene close to the charge-neutrality point has been predicted if the momentum of the SAW exceeds that of the graphene charge carriers at the Fermi level. Moreover, by the application of a DC current to the graphene layer, SAW amplification has been observed experimentally when the drift velocity of the graphene charge carriers exceeded the velocity of the SAW. 

The use of SAWs for the modulation of the graphene electronic properties has also been theoretically investigated. Due to the slow propagation velocity of the SAW with respect to the Fermi velocity of the charge carriers in graphene, the periodic potentials of the SAW have been proposed as superlattices for the modification of the electronic energy dispersion. In addition, it has been predicted that the periodic elastic deformations of SAWs can act as a diffraction grating for the efficient coupling of light to plasmons in graphene. 

The effect of the interaction between SAWs and graphene that has been most widely  studied experimentally until now is the generation of acousto-electric currents. The amplitude of the AE depends linearly on the power and frequency of the SAW but, in contrast to conventional DC currents, the AE current does not increase monotonously with the conductivity of the electron gas: it becomes inversely proportional to the conductivity above a characteristic value that depends on the SAW velocity and on the dielectric properties of the materials surrounding the graphene layer. In addition, the AE current reverses its sign when the graphene carrier polarity changes from electrons to holes. These properties are, in general, well described by the classical relaxation model between SAWs and a 2D electron gas. At low temperatures and SAW frequencies, however, the behavior of the AE current measured in CVD graphene transferred to a piezoelectric substrate differs from the predictions of the relaxation model due to the contribution of defects and charged impurities to the graphene electronic conductivity over the wavelength and frequency probed by the SAW.

The interesting functionalities derived from the interaction between SAWs and graphene have mostly been demonstrated in graphene layers transferred to the surface of piezoelectric insulators like LiNbO$_3$ or LiTaO$_3$. SAW-induced acousto-electric currents have also been reported in epitaxial graphene on SiC. This approach is interesting for technological applications due to the availability of large area graphene layers directly on the surface of a semi-insulating wafer. The weak piezoelectricity of SiC, however, makes it necessary to use strong piezoelectric materials for the efficient generation of SAWs and their coupling to the graphene charge carriers. To this end, the coating of graphene on SiC with a piezoelectric ZnO film has allowed the observation of AE currents that are thousand times larger than those induced in non-coated graphene, where the acousto-electric interaction was mediated only by the piezoelectricity of SiC.

The demonstration of some of the unique properties of graphene discussed in this topical review require electronic excitations moving ballistically along the graphene layer. In the case of the interaction between graphene and SAWs, this means that the mean-free-path of the graphene charge carriers must exceed the SAW wavelength, which is typically of the order of a few micrometers. Due to the single-layer nature of graphene, its electronic properties are strongly influenced by the materials below and above it, thus drastically diminishing its electron mobility. Until now, the most successful strategy to screen the deleterious influence of the surroundings has been the encapsulation of graphene in hexagonal boron nitride (h-BN) films, where mean-free-paths as large as 28~$\mu$m have been reported~\cite{Banszerus_NL16_1387_16}. A first evidence of the benefits of the incorporation of h-BN in graphene acousto-electric devices has been recently provided by Fandan~\etal~\cite{Fandan_JPhysD51_204004_2018}, where they predict that the intercalation of an h-BN film between the graphene and the piezoelectric AlN substrate strongly enhances the SAW-mediated coupling between light and graphene plasmons.

While the realization of encapsulated graphene relies currently on stacked flakes fabricated by transfer techniques~\cite{Mayorov_NanoLetters11_2396_2011, Banszerus_NL16_1387_16}, recent advances in the direct growth of graphene/h-BN heterostructures using molecular beam epitaxy~\cite{Zuo_SciRep5_14760_2015, Wofford_SciRep7_43644_2017} suggest the future availability of on-demand, large-area graphene/h-BN multilayers with high crystalline quality. It is therefore reasonable to expect that graphene encapsulated in h-BN will be the platform for the experimental demonstration of SAW functionalities that, until now, have just been predicted theoretically. On the other hand, the recent covering of epitaxial graphene by a thin MgO interlayer while preserving its electronic properties could open the way to the use of other materials than h-BN for the protection of the graphene electronic excitations from the perturbations induced by the neighboring layers.

Although this topical review has focused on the interaction of SAWs with graphene, it is important to remark that acousto-electric modulation has also been reported in other two-dimensional materials like transition metal dichalcogenides~\cite{Preciado_NatComm6_8593_2015, Huang_JPhysD50_114005_2017} and black phosphorous~\cite{Zheng_IEEETransNanotech17_590_2018}. Moreover, the demonstration of the coupling of SAWs to van der Waals p-n diodes consisting of stacked black phosphorous and MoS$_2$ films~\cite{Zheng_Nanoscale10_10148_2018} opens very promising perspectives for the future acoustic manipulation of electronic excitations in e.g. field-effect transistors consisting exclusively of the combination of different two-dimensional materials~\cite{Roy_ACSNano8_6259_2014}.

Finally, graphene and other two-dimensional materials have also been incorporated in optical structures like photonic crystal nanocavities and optical modulators~\cite{Liu_Nature474_64_2011, Gan_APL103_181119_2013, Gao_NanoLetters15_2001_2015, Hammer_SciRep7_7251_2017}. As the performance of these optical systems can be controlled using acoustic fields~\cite{Fuhrmann_NP5_605_11}, future developments in the manipulation of two-dimensional materials by SAWs will probably include their acousto-optic modulation by means of such structures. In addition, the discovery of luminescent centers in WSe$_2$ and h-BN acting as single photons sources~\cite{Srivastava_NatNanotech10_491_2015, Tran_NatNano11_37_2016} opens the way to the acoustic manipulation of single photon emitters in 2D materials in a similar way as for self-assembled quantum dots~\cite{Metcalfe_PRL105_37401_10} and nitrogen-vacancy centers in diamond~\cite{Golter_PRL116_143602_16}. Furthermore, the recent readout of the quantum states of defects in diamond by means of the electric currents generated in an adjacent graphene film~\cite{Brenneis_NN10_135_15} suggests that the combined acousto-electric and acousto-optic modulation of heterostructures consisting of two-dimensional materials could be a promising platform for the development of new functionalities for quantum opto-electronics and quantum communication.

\section*{Acknowledgments}
The authors acknowledge Manfred Ramsteiner for discussions. This publication is part of a project that has received funding from the European Union's Horizon 2020 research and innovation programme under the Marie Sk\l{}odowska-Curie grant agreement No 642688. A.H.M. also acknowledges financial support by the Deutsche Forschungsgemeinschaft (Project HE 6985/1-1 within the Priority Programme SPP 1459 Graphene).

\section{References}

%Bibliographystyle
\bibliographystyle{unsrt}
%\bibliography{Graphene,literature,mypapers}

\end{document}